\documentclass[aps,pre,a4paper,10pt,twocolumn,notitlepage,footinbib,superscriptaddress,longbibliography]{revtex4-1}
\usepackage[english]{babel}
\usepackage{lmodern}
\usepackage{endnotes}
\usepackage{amssymb,amsmath,amsfonts}
\usepackage{textcomp}
\usepackage{url}
\usepackage{subfigure}
\usepackage{soul}

\usepackage{graphicx}
\usepackage{dcolumn}
\usepackage{bm}
\usepackage{xcolor}

\begin{document}

\title{Lattice Boltzmann modeling of cholesteric liquid crystal droplets under an oscillatory electric field}

\author{F. Fadda}
\affiliation{Institute of Physics, University of Amsterdam, 1098 XH Amsterdam, The Netherlands.}
\affiliation{Department of Chemical Engineering, Kyoto University, Kyoto 615-8510, Japan.}
\affiliation{Dipartimento di Fisica and Sezione INFN, Università di Bari, Via Amendola 173, 70126 Bari, Italy.}
\author{A. Lamura}
\affiliation{Istituto Applicazioni Calcolo, CNR, Via Amendola 122/D, 70126 Bari, Italy.}
\author{A. Tiribocchi}
\affiliation{Istituto Applicazioni Calcolo CNR, via dei Taurini 19, 00185 Rome, Italy.}

\date{\today}

\begin{abstract}
We numerically study the dynamics of quasi-two dimensional cholesteric liquid crystal droplets  in the presence of a time-dependent electric field, rotating at constant angular velocity. A surfactant sitting at droplet interface is also introduced to prevent droplet coalescence. The dynamics is modeled following a hybrid numerical approach, where a standard lattice Boltzmann technique solves the Navier-Stokes equation and a finite difference scheme integrates the evolution equations of liquid crystal and surfactant. Our results show that, once the field is turned on, the liquid crystal rotates coherently triggering a concurrent orbital motion of both droplets around each other, an effect due to the momentum transfer to the surrounding fluid. In addition the topological defects, resulting from the conflict orientation of the liquid crystal within the drops, exhibit a chaotic-like motion in cholesterics with a high pitch, in contrast with a regular one occurring along circular trajectories observed in nematics drops. Such behavior is found to depend on  magnitude and frequency of the applied field as well as on the anchoring of the liquid crystal at the droplet interface. These findings are quantitatively evaluated by measuring the angular velocity of fluid and drops for various frequencies of the applied field.
\end{abstract}

\maketitle

\section{Introduction}
Nematic liquid crystals are an example of soft material in which the local alignment of anisotropic-shaped molecules they are made of is described by a unit magnitude director field $\mathbf{n}$ with head-tail symmetry. Cholesteric liquid crystals, on the contrary, are chiral systems in which the locally favoured state of the director field is a twist deformation in the direction perpendicular to the molecules~\cite{degennes,chandra,oswald}. Such helical arrangement is characterized by a helix pitch $p_0$, a quantity measuring the distance over which the director rotates by $2\pi$.

Of particular relevance to us are cholesteric liquid crystal droplets, highly confined chiral soft fluids that have found  vast application in several sectors of modern industry, ranging from photonics \cite{EXP7,EXP8} and laser beams \cite{EXP4} to microlasers \cite{EXP3}, optics \cite{EXP6}, displays \cite{noh2014} and, more recently, as active material \cite{LC7,LC10,carenza}. In these objects the order of the director is crucially affected by the anchoring of the liquid crystal at the droplet interface \cite{lopez,chol00,zumer,zumer2,LC13,orlova,car_prl}. Indeed, under confinement, the typical helical structure of the cholesteric may conflict with that imposed at the boundaries, often favouring the formation of topological defects (or disclinations) whose nature can decisively condition  mechanical and optical properties of the liquid crystal \cite{zumer4,zumer,Li_natcomm}.
While, over the years, considerable efforts have been addressed to theoretically investigate the physics of cholesteric droplets and their associated defect structure at equilibrium \cite{zumer,zumer2,copar,darmon_2016,LC15,LC16,tran_2017,cuboidal,chol0}, only recently a number of numerical works have been dedicated to pinpointing their response under an external driving, such as a heat flux \cite{yoshioka_2018} or an electric field \cite{fadda,fadda3}. Such works have been inspired by experiments showing for example that, if subject to a temperature gradient, cholesteric drops are set into rotation due to either a thermomechanical torque mechanism \cite{lehmann,oswaldprl,oswaldepje} or to Marangoni flows  \cite{yoshioka_2018,oswald_soft}. A rotation can be alternatively triggered by applying a uniform (i.e. time independent) and large enough electric field, giving rise to a torque applied to the liquid crystal confined within the drops \cite{madhu_1987,madhu_1989,tarasov2003,skaldin2018}. Further experiments have also shown that angular velocity and shape of such rotating drops can be controlled by tuning an oscillatory electric field coupled to the liquid crystal subject to a thermal gradient \cite{lehmann2}.

In a previous work \cite{fadda} we numerically studied the response of a quasi-two dimensional cholesteric drop dispersed in an isotropic fluid solely subject to an electric field coupled to the liquid crystal, and we showed that its dynamics and that of the defects critically depend on magnitude and direction of the field as well as on elasticity and pitch of the liquid crystal. If the field is non uniform, such as a rotating one with constant frequency, the defects display a persistent periodic motion occurring with an angular speed generally lower than that set by the field, due to the anisotropy of the liquid crystal. In this work we go one step further and  consider a couple of cholesteric drops in an isotropic fluid subject to a time-dependent electric field ${\bf E}({\bf r},t)$ rotating at constant frequency $\omega$. Droplet coalescence is prevented by including a surfactant accumulated at their interfaces.
The theoretical framework used to describe the droplet physics relies on well-established continuum prescriptions \cite{mazur}, in which a small number of continuum fields, such as concentration and ordering of the liquid crystal, amount of surfactant, density and velocity of the fluid, capture the coarse-grained behavior of the system. The evolution of these fields is written in terms of a set of hydrodynamic equations in which the thermodynamic forces (such as pressure tensor and molecular field) stem from functional differentiation of a free-energy encoding the equilibrium properties. 
Following previous studies \cite{dennis,henrich2,sulaiman}, we simulate the droplet dynamics using a hybrid lattice Boltzmann (LB) formulation \cite{succi}, in which the Navier-Stokes equation governing the evolution of the fluid velocity is solved using a standard LB approach, while advection-relaxation equations of liquid crystal and surfactant are integrated using a finite-difference scheme.

Our results report a complex scenario in which, regardless of the pitch of the cholesterics and of its anchoring at the droplet interface (perpendicular or tangential), the rotation of the liquid crystal triggered by the applied field fosters that of the fluid confined within and in the surroundings of the drops, an effect overall akin to the dynamics observed in \cite{fadda}. However, in a double-drop configuration the fluid also favours the rotation of both drops around an axis located in the fluid film separating the droplets. Their angular velocity as well as that of the fluid confined within increase approximately linearly  for low values of frequency of the applied field while, for higher ones, it diminishes and stabilizes to constant values. 
This behavior affects the defect dynamics too. While in the nematic limit (i.e. infinite pitch) topological defects of charge $\pm 1/2$ follow a circular path  either close to the interface or towards the center of the drops, in the cholesteric phase $\lambda$ and $\tau$ defects emerge \cite{zumer4} (see also section III A), the latter firmly anchored to the former which periodically stretch and shorten under the oscillatory field.

The paper is organized as follows. In Section 2 we describe the thermodynamics of a cholesteric droplet hosted in an isotropic medium and the numerical implementation of the computational model, while in Section 3 we illustrate the results. In particular, we start off with studying the dynamic response of two nematic drops under an oscillatory electric field and then we consider two cholesteric drops for various interface anchoring conditions and frequency of the field. Some final remarks close the manuscript.

\section{Model}

We consider two cholesteric droplets immersed in an isotropic host in the presence of a surfactant absorbed onto their interface. The physics of such system is described by a set of coarse-grained fields, $\rho(\textbf{r},t)$, $\phi(\textbf{r},t)$, $c(\textbf{r},t)$, $\textbf{u}(\textbf{r},t)$ and $Q_{\alpha\beta}(\textbf{r},t)$ which represent, respectively, the mass density, the concentration of the cholesteric phase relative to the isotropic one, the concentration of the surfactant, the average velocity of the fluid, and the tensor order parameter that, within the Beris-Edwards framework \cite{beris,chandra,degennes}, captures the ordering of the liquid crystal. In the uniaxial approximation, $Q_{\alpha\beta}=q(n_{\alpha}n_{\beta}-\frac{1}{3}\delta_{\alpha\beta})$ (Greek subscripts denote the Cartesian coordinates), where $\textbf{n}$ is the director field accounting for the local direction of the molecules and $q$ gauges the amount of the local order which is proportional to the largest eigenvalue of $\textbf{Q}$ ($0\leq q \leq \frac{2}{3}$).

The equilibrium properties of this system in the presence of an external electric field are described by a Landau-de Gennes free energy $\mathcal{F}=\int_{V}fdV$, where the free-energy density is
\begin{eqnarray}\label{free_E}
f&=&\frac{a}{4}\phi^{2}(\phi-\phi_{0})^{2}+\frac{(\kappa_{\phi}+\kappa_{c}c)}{2}(\nabla \phi)^2+c \ln c\nonumber\\&&+A_0 \biggr[\frac{1}{2}\left(1-\frac{\zeta(\phi)}{3}\right)Q^2_{\alpha\beta}-\frac{\zeta(\phi)}{3}Q_{\alpha\beta}Q_{\beta\gamma}Q_{\gamma\alpha}\nonumber\\
&&+ \frac{\zeta(\phi)}{4}(Q_{\alpha\beta}^2)^2\biggl]\nonumber\\
&&+ \frac{K}{2}\biggl[(\partial_{\beta}Q_{\alpha\beta})^2+(\varepsilon_{\alpha\gamma\delta}\partial_{\gamma}Q_{\delta\beta}+ 2q_0Q_{\alpha\beta})^2\biggr]\nonumber\\
&&+W(\partial_{\alpha}\phi)Q_{\alpha\beta}(\partial_{\beta}\phi)-\frac{\epsilon_{a}}{12\pi}E_{\alpha}Q_{\alpha\beta}E_{\beta}.
\end{eqnarray}

The first term, multiplied by the positive constant $a$, is the binary fluid bulk free energy which ensures the existence of two coexisting minima, $\phi=\phi_0$ within the droplet (where the cholesteric liquid crystal is confined) and $\phi=0$ outside. The second term of Eq.~(\ref{free_E}) describes the interfacial properties of the mixture. The constant $\kappa_{\phi}$ controls surface tension and interface width which, in a binary fluid without liquid crystal, are
$\sigma\sim \sqrt{ak_{\phi}}$  and $\xi\sim \sqrt{k_{\phi}/a}$ respectively, while $\kappa_c$ determines whether the surfactant accumulates either at the droplet interface ($\kappa_c<0$) or in the droplet bulk ($\kappa_c>0$). Throughout our simulations $\kappa_c$ is kept negative \cite{lamura1,lamura2,fadda,fadda2}. The logarithmic term, $c\ln c$, stems from the translational entropy of the surfactant \cite{Yoshinaga}.

The bulk properties of the liquid crystal are captured by three further contributions (where summation over repeated indices is assumed) which contain terms of the $Q$-tensor up to the fourth order. The scale factor $A_0$ is a positive constant while $\zeta(\phi)$ controls the isotropic-liquid crystal transition and determines which of the two phases is stable. For a nematogen without chirality ($q_0=0$), the global minimum of the free energy is the nematic state for $\zeta(\phi)\geq\zeta_c=2.7$ and the isotropic one for $\zeta(\phi)<\zeta_c$. Following previous works \cite{sulaiman,fadda}, we set $\zeta=\zeta_0+\zeta_s\phi$, where $\zeta_0$ and $\zeta_s$ control the boundary of the coexistence region. Local distortions of the liquid crystal enter the free energy through first order gradient terms of ${\bf Q}$ augmented by a gradient-free contribution which guarantees that the free energy is positive. $K$ is the elastic constant, $\varepsilon_{\alpha\gamma\delta}$ is the Levi-Civita antisymmetric tensor and $q_0=2\pi/p_0$ is the chirality which sets the pitch length $p_0$ of the cholesteric.        
The anchoring of the director field at the droplet interface is ensured by the term proportional to $W$ whose value gauges the anchoring strength. In our simulations we are in the strong anchoring regime, meaning that the director field at the interface is only weakly affected by an external field. The sign of $W$ controls the orientation of the liquid crystal: if positive, the director is aligned tangentially to the interface (planar or tangential anchoring) whereas, if negative, it is aligned perpendicularly (homeotropic anchoring).
Finally, the last term of the free energy accounts for the interaction between the liquid crystal and the external electric field $E$, where $\epsilon_{\alpha}>0$ is the dielectric anisotropy.

It is often convenient to write the free energy in terms of a decreased number of dimensionless parameters on which the phase behavior can depend \cite{mermin}:
\begin{equation}
\kappa=\sqrt{\frac{108Kq_{0}^{2}}{A_{0}\zeta(\phi_{0})}},
\end{equation}
\begin{equation}
\tau=\frac{27(1-\zeta(\phi_{0})/3)}{\zeta(\phi_{0})},
\end{equation}
\begin{equation}
\varepsilon^{2}=\frac{27\epsilon_{a}}{32\pi A_{0}\zeta(\phi_{0})}E_{\alpha}E_{\alpha}.
\end{equation}
Here $\tau$ is the reduced temperature, multiplying the quadratic terms of the dimensionless bulk free energy, $\kappa$ is the chirality multiplying the gradient ones and $\varepsilon$ is an effective field strength. 

The dynamic equations governing the evolution of the system are \cite{mazur,beris}
\begin{equation}
\label{phi}
\partial_{t}\phi+{\bf u}\cdot \nabla\phi
=M \nabla^{2} \mu_{\phi},
\end{equation}
\begin{equation}\
\label{c}
\partial_{t}c+{\bf u}\cdot\nabla c=\nabla \cdot \left [ L(c)\nabla \mu_c \right ],
\end{equation}
\begin{equation}\label{Q}
(\partial _{t}+\textbf{u}\cdot \nabla)\mathbf{Q}-\textbf{S}(\mathbf{W},\mathbf{Q})=\Gamma \mathbf{H},
\end{equation}
\begin{equation}\label{cont_eq}
\nabla \cdot \textbf{u}=0,
\end{equation}
\begin{equation}\label{nav_stok}
\rho(\partial_{t}+u_{\beta}\partial_{\beta})u_{\alpha}=\partial_{\beta}P_{\alpha \beta}.
\end{equation}
The first two equations describe the dynamics of the $\phi$ and $c$ fields. In Eq.~(\ref{phi}) $\mu_{\phi}=\delta\mathcal{F}/\delta \phi$ is the chemical potential and $M$ is the mobility, while in Eq.~(\ref{phi}) $\mu_c=\delta\mathcal{F}/\delta c$ and $L(c)=Dc$, where $D$ is the diffusion coefficient of the surfactant. The functional form of $L(c)$ is necessary to avoid a singularity at $c=0$ in the surfactant density current ${\bf j}_c= - L(c)\nabla\mu_c$ \cite{Yoshinaga,fadda2}. The dynamics of ${\bf Q}$ is described by Eq.~(\ref{Q}), where the term on the left hand side is a generalized material derivative. In particular $\textbf{S}(\textbf{W},\textbf{Q})$ takes into account the fact that the order parameter distribution can be rotated or stretched by the flow, and can be written as \cite{beris}
\begin{eqnarray}
\mathbf{S}(\mathbf{W},\mathbf{Q})&=&(\xi \mathbf{D}+\boldsymbol{\Omega})(\mathbf{Q}+\mathbf{I}/3)+(\mathbf{Q}+\mathbf{I}/3)(\xi \mathbf{D }- \boldsymbol{\Omega})\nonumber\\&& -2\xi(\mathbf{Q}+\mathbf{I}/3)Tr(\mathbf{Q}\mathbf{W}).
\end{eqnarray}
Here $\mathbf{D}=(\mathbf{W}+\mathbf{W}^{T})/2$ and $\mathbf{\Omega}=(\mathbf{W}-\mathbf{W}^{T})/2$ are the symmetric and antisymmetric part of the tensor gradient velocity $W_{\alpha\beta}=\partial_{\beta}u_{\alpha}$ and $\textbf{I}$ is the identity matrix. The constant $\xi$ determines the aspect ratio of molecules; if positive,  molecules are rod shaped while, if negative, they are disk-like.
It also controls the response of a nematic liquid crystal under shear flow. Indeed, at the steady state (achieved, for example, after imposing a homogeneous shear), the director aligns with the flow gradient at an angle $\theta$ fulfilling the relation $\xi cos(2\theta)=(3q)/(2+q)$ \cite{degennes}. Real solutions, corresponding to a flow aligning regime, are obtained if $\xi\geq 0.6$.
Finally, $\Gamma$ is the collective rotational 
diffusion constant which, together $q$ (the scalar order parameter), controls the rotational viscosity $\gamma_1=2q^2/\Gamma$ of the liquid crystal explicitly appearing in the Leslie-Ericksen theory of nematodynamics \cite{degennes,dennis}.  In Eq.\ref{Q},  $\textbf{H}$ is the molecular field which is given by
\begin{equation}
\mathbf{H}=-\frac{\delta \mathcal{F}}{\delta \mathbf{Q}}+\frac{\mathbf{I}}{3}Tr \frac{\delta \mathcal{F}}{\delta \mathbf{Q}}.
\end{equation}
The last two equations are the continuity and the Navier-Stokes equation (in the incompressible limit), where $P_{\alpha \beta}$ is the total stress tensor given by
\begin{equation}\label{stress}
P_{\alpha\beta}=s_{\alpha\beta}+\pi_{\alpha\beta}.
\end{equation}
In Eq.~(\ref{stress}) $s_{\alpha\beta}=\eta(\partial_{\alpha}u_{\beta}+\partial_{\beta}u_{\alpha})$  is the stress of the background fluid with $\eta$ isotropic shear viscosity, while $\pi_{\alpha\beta}$ 
can be written as the sum of three further terms \cite{beris,degennes,henrich2,sulaiman}  
\begin{equation}
\pi_{\alpha\beta}=\sigma_{\alpha\beta}+\tau_{\alpha\beta}+\Pi_{\alpha\beta},
\end{equation}
where
\begin{eqnarray}\label{symm}
\sigma_{\alpha \beta}&=&-P\delta_{\alpha\beta}-\xi H_{\alpha\gamma}\left ( Q_{\gamma\beta}+\frac{1}{3}\delta_{\gamma\beta} \right )\nonumber\\&&-\xi\left ( Q_{\alpha\gamma}+\frac{1}{3}\delta_{\alpha\gamma} \right )H_{\gamma\beta}\nonumber\\&&
 +2\xi\left ( Q_{\alpha\beta} -\frac{1}{3}\delta_{\alpha\beta}\right)Q_{\gamma\mu}H_{\gamma\mu},
\end{eqnarray}
\begin{equation}\label{antsymm}
\tau_{\alpha\beta}=Q_{\alpha\nu}H_{\nu\beta}-H_{\alpha\nu}Q_{\nu\beta},
\end{equation}
and
\begin{eqnarray}\label{int_st}
\Pi_{\alpha\beta}&=&-\left ( \frac{\delta \mathcal{F}}{\delta \phi}\phi + \frac{\delta{\cal F}}{\delta c}c -\mathcal{F}\right )\delta_{\alpha\beta}-\frac{\delta \mathcal{F}}{\delta (\partial_{\beta}\phi)}\partial_{\alpha}\phi\nonumber\\&&-\frac{\delta \mathcal{F}}{\delta (\partial_{\beta}Q_{\gamma\mu})}\partial_{\alpha}Q_{\gamma\mu},
\end{eqnarray}
with $P$ isotropic pressure. Here $\sigma_{\alpha\beta}$ and $\tau_{\alpha\beta}$ are the symmetric and antisymmetric part of the liquid crystal stress tensor, while $\Pi_{\alpha\beta}$ includes stress contributions of binary fluid and surfactant plus interfacial terms.

\subsection{Numerical implementation}
Eqs.~(\ref{phi})-(\ref{nav_stok}) are solved by using a hybrid numerical approach, in which Eqs.~(\ref{phi}), (\ref{c}) and (\ref{Q}) are integrated by means of a finite difference-scheme while Eqs.~(\ref{cont_eq}) and (\ref{nav_stok}) by a standard lattice Boltzmann method \cite{succi}. We incidentally note that, although a hybrid LB machinery has been already successfully employed to study binary \cite{tiribocchi} and ternary fluids \cite{fadda2}, active gels \cite{act1,sm19,sr20} and liquid crystals, such as nematics \cite{tiribocchi2,chol2} and cholesterics \cite{tiribocchi4,fadda,fadda3,chol1,chol3,chol4}, in this work the applicability of the method has been further extended to study liquid crystal emulsions in the presence of a surfactant whose dynamics is explicitly solved.

Here we shortly outline the details of the computational model. The lattice Boltzmann method is built starting from a set of distribution functions $f_{i}(\textbf{x},t)$ (defined on a lattice site ${\bf x}$ at time $t$) whose sum on each site ${\bf x}$ gives the density $\rho$ of the fluid. The $f_i$ evolve following a discrete Boltzmann equation
\begin{eqnarray}\label{fi}
f_{i}(\textbf{x}+\textbf{e}_{i}\Delta t, t+\Delta t)&=&f_{i}(\textbf{x},t)+
\frac{\Delta t}{2}[\mathcal C_{f_{i}}(\textbf{x},t,\{f_{i}\})\nonumber\\&&+\mathcal C_{f_{i}}(\textbf{x}+\textbf{e}_{i}\Delta t, t+\Delta t, \{f_i^*\})],\nonumber\\
\end{eqnarray}
where $\Delta t$ is the integration time-step and ${\bf e}_i$ are velocity vectors linking nearest neighbor sites. In our simulations we employ a $D3Q15$ scheme, i.e. a cubic lattice with $15$ velocity vectors $\mathbf{e}_{i}^{(0)}=(0,0,0)$, $\mathbf{e}_{i}^{(1)}=(\pm 1,0,0), (0,\pm 1, 0), (0,0,\pm 1)$ and $\mathbf{e}_{i}^{(2)}=(\pm 1,\pm 1 ,\pm 1)$, where
the index $i$ runs from $0$ to $14$ and is defined so that $i=0$ corresponds to $\mathbf{e}_{i}^{(0)}$, $i=1,...,6$ to $\mathbf{e}_{i}^{(1)}$ (nearest neighbors) and $i=7,...,14$ to $\mathbf{e}_{i}^{(2)}$ (next-nearest neighbors) \cite{dennis}. In Eq.(\ref{fi}), the $f_{i}^{*}$ are a first order approximation to $f_{i}(\textbf{x}+\textbf{e}_{i}\Delta t, t+\Delta t)$ and are obtained by applying $\Delta t\mathcal C_{f_{i}}(\textbf{x},t,\{f_{i}\})$ on the right hand side of Eq.~(\ref{fi}). Such approach, analogous to a predictor-corrector scheme, has been proven to enhance  the numerical stability of the method \cite{dennis,henrich2}. 
The term $\mathcal{C}_{f_i}$ is the collision operator, which is given by 
\begin{equation}
\mathcal C_{f_{i}}(\textbf{x},t,\{f_{i}\})=-\frac{1}{\tau}(f_{i}(\textbf{x},t)-f_{i}^{eq}(\textbf{x},t,\{f_{i}\})) +p_{i}(\mathbf{x},t,\{f_{i}\}),
\end{equation}
where $\tau$ is a relaxation time (controlling the fluid viscosity $\eta=\rho\tau/3$ \cite{dennis,henrich2}), $f_i^{eq}$ are the equilibrium distribution functions which can be written as a second order expansion in the fluid velocity ${\bf u}$ and  the terms $p_i$ represent driving contributions. Following previous works \cite{wagner,sulaiman,henrich2}, a considerable reduction of the spurious velocities (non-zero velocity field at equilibrium caused by different discretisation of pressure tensor and molecular field) is ensured if $\sigma_{\alpha\beta}$ enters the second moment of $f_i^{eq}$ while $\tau_{\alpha\beta}$ and $\Pi_{\alpha\beta}$ enter the first moments of $p_i$. However, as previously mentioned, our method is different from the ones in Refs.\cite{henrich2} and \cite{sulaiman},  since it solves the equations of $\phi$ and $c$, not included in \cite{henrich2}, using a hybrid approach, not adopted in \cite{sulaiman}. Finally, imposing conservation of mass and momentum to $f_i^{eq}$, the continuity and the Navier-Stokes equations are recovered via a Chapman-Enskog expansion of Eq.(\ref{fi}).

Unlike Equations (\ref{cont_eq}) and (\ref{nav_stok}),
Equations (\ref{phi}), (\ref{c}) and (\ref{Q}) are solved using a predictor-corrector scheme,  in which the finite difference operators
(spatial derivatives and laplacian) are discretised using a stencil representation and the integration time-step is set equal to that of the lattice Boltzmann \cite{tiribocchi}. The advantage of using a hybrid LB method with respect to a full LB approach (such as the one used in Refs.\cite{sulaiman,dennis}) is that it allows simulations of large systems with substantially smaller memory requirements. Indeed, a full LB treatment of our cholesteric droplet would require to store eight sets of fifteen distribution functions (if a $D3Q15$ scheme is used), i.e. one set for the fluid, five sets for the liquid crystal (since the ${\bf Q}$ tensor has five independent components), one for $\phi$ and a further one for $c$. On the contrary, the hybrid method needs one set of fifteen $f_i$ for the fluid plus seven independent components for the remaining fields.

\subsection{Initial conditions, parameter values and mapping to physical units}

Simulations are performed on a quasi-2d squared lattice ($L_x=1$, $L_{y}=250, L_{z}=250$), periodic in all directions, in which either an isolated or a couple of cholesteric drops of equal size are surrounded by an isotropic fluid. In the former system the droplet is placed in the center of the mesh while in the second case the centers of mass of the drops are initially located at distance $d>2R$, where $R$ is the radius of each drop kept fixed to $32$ lattice sites. With respect to a fully 3d study, this quasi-2d setup allows for simulations at a much reduced computational cost and concurrently preserves the intrinsic three-dimensional structure of the cholesteric liquid crystal, since it enables out-of-plane components of the macroscopic fields (along the $x$-direction).

The system is initialized as follows. We have set $\phi=0$ and ${\bf Q}=0$ outside the droplets, while inside $\phi$ is kept constant (equal to $\phi_0=2$) and the components of ${\bf Q}$ are chosen to accommodate a cholesteric with helix parallel to the horizontal $y$-axis.  This is achieved by setting 
\begin{equation}
Q_{xx} =  (b_{0} - b_{1}/ 2)cos(2q_{0} y) + b_{1} / 2,
\end{equation}
\begin{equation}
Q_{yy} = - b_{1},
\end{equation}
\begin{equation}
Q_{xz} =  - (b_{0} - b_{1}/ 2)sin(2q_{0}y) 
\end{equation}
\begin{equation}
Q_{xy} = Q_{yz}=0,
\end{equation}
where $b_{0} = 0.546$, $b_{1} = 0.272$.
The parameter $q_0=2\pi/p_0$ controls the number $N$ of $\pi$ twists of the liquid crystal in a droplet of diameter $2R$. Indeed, to compare $p_0$ with the size of the droplet, one can define the pitch length as $p_0=4R/N$. If, for instance, $R=32$ and $q_{0}=2\pi/32$, one has $N=4$, while, if $q_{0}=2\pi/64$, $N=2$. In particular, $N=0$ corresponds to a nematic liquid crystal. 
Finally the concentration of the surfactant $c$ is initially set to a constant value equal to $c_0=0.02$ uniformly on the lattice.

In our simulations the numerical values of the parameters are $a=0.07$, $\kappa_{\phi}=0.14$, $M=0.05$, $\kappa_c=-1.3$, $D=0.1$, $A_0=1$, $K=0.03$, $|W|=0.04$, $\xi=0.7$, $\Gamma=1$, $\eta=1.67$ and $\epsilon_a\simeq 10$. Also, lattice spacing and integration time step are $\Delta x=1$ and $\Delta t=1$. Finally, by setting $\zeta(\phi_0)=3$ and $q_0=\{2\pi/32,2\pi/64\}$, one has $\tau=0$ and $\kappa\simeq{0.1,0.2}$, well within the region of the cholesteric phase according to the phase diagram of Ref. \cite{dupuis,henrich}. 

An approximate mapping to real physical values can be obtained by assuming that one space, time and force in LB units correspond to $L=10^{-1}\mu$m, $T=1\mu$s and $F=1$nN. This corresponds to a cholesteric liquid crystal of elastic constant $\sim 30$pN and rotational viscosity of $\sim 1$ Poise confined within a droplet of diameter $\sim 10\mu$m. The mapping of the electric field to real units  can be performed by using the dimensionless parameter $\varepsilon$. As in previous works \cite{fadda,tiribocchi4}, by assuming $\frac{27}{2A_{0}\zeta(\phi_{0})}\approx 1-5\times 10^{-5}J^{-1}m^{-3}$, an electric field of $1-10 V/\mu m$  with dielectric constant of the order $10$ (a positive value ensures that the director field aligns parallel to the electric field), gives $\varepsilon^{2}\approx 0.001$. This value ensures that non-local effects associated with inhomogeneities of the electric field can be negnected \cite{shiya}. Also, the values
of the frequency $\omega$ of the applied rotating field  typically vary from $10^{-5}$ (low frequency regime) to
$10^{-1}$ (high frequency regime). If, for example, $\omega \simeq 10^{-2}$, the director field within the droplet would complete a full turn in approximately 600 simulation time steps, which
would roughly correspond to $0.5$ms in real time.
Finally, the anchoring constant is mapped to a value $W\approx10^{-4}Jm^{-2}$ \cite{sulaiman,anderson,fadda}.

\section{Results}

We start by discussing the equilibrium configurations of an isolated cholesteric droplet and afterwards we focus on the dynamics of couple of droplets under a  rotating (or oscillatory) electric field.

\subsection{Equilibrium states}

In Ref.\cite{fadda} we have extensively described the typical equilibrium configurations of a quasi-2d isolated cholesteric droplet, whose liquid crystal profile was found to essentially depends upon elasticity, direction of surface anchoring (perpendicular or tangential) and number $N$ of $\pi$ twists of the cholesteric. Here we shortly recap the essential features.

\begin{figure*}
\includegraphics[width=1.\linewidth]{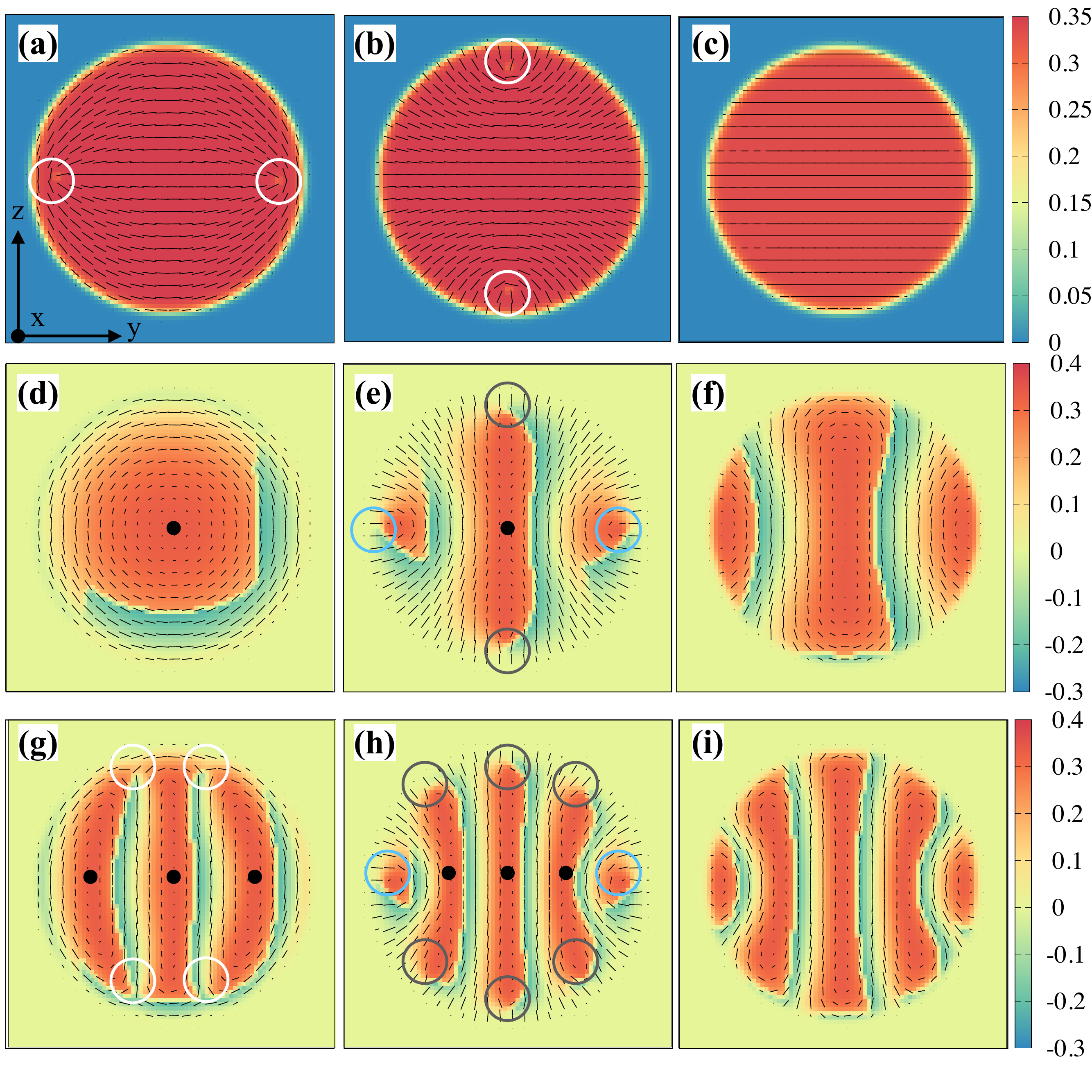}
\caption{Equilibrium configurations of nematic and cholesteric liquid crystal droplets in an isotropic fluid. The radius is $R=32$, the anchoring strengths are $W=0.04$ (a,d,g), $W=-0.04$ (b,e,h) and $W=0$ (c,f,i), while the number of $\pi$ twists is  $N=0$ (a,b,c), $N=2$ (d,e,f) and $N=4$ (g,h,i). Topological defects, where the orientational order drops, are highlighted with circles; the white ones indicate generic in-plane defects, the grey ones twist-disclinations and the blue ones $\tau$ defects. The black spot marks the lambda regions. The colour map of (a)-(b)-(c) (nematic droplets) shows the largest eigenvalue of the $\boldsymbol{Q}$-tensor and ranges from $0$ (blue, the isotropic region) to $\simeq 0.33$ (red, the liquid crystal). The color map of (d)-(e)-(f)-(g)-(h)-(i) (cholesteric droplets) shows the x-component of the director multiplied by the largest eigenvalue of the $\boldsymbol{Q}$-tensor. This is to highlight regions where the director field has components out of the $y-z$ plane. The red zone, in particular, marks the exit from the plane towards the reader.}
\label{fig2}
\end{figure*}

In Fig.~\ref{fig2} we show a number of selected cases observed for $N=0$ (a,b,c), $N=2$ (d,e,f) and $N=4$ (g,h,i), where $W=0.04$ (tangential direction) in Fig.~\ref{fig2}a,d,g, $W=-0.04$ (perpendicular direction) in Fig.~\ref{fig2}b,e,h and $W=0$ (no fixed anchoring) in Fig.~\ref{fig2}c,f,i with no surfactant (its presence does not alter the equilibrium states). 

If $N=0$ (i.e. a nematic state) and $W\neq 0$, two topological defects (highlighted with white circles), placed on opposite sides and near the interface, emerge due to the conflict orientation between the direction of the liquid crystal in the bulk and that at the edge. In particular, for both tangential and perpendicular anchoring (Fig.~\ref{fig2}a-b), the topological charge is $1/2$. On the contrary, if $W=0$ (Fig.~\ref{fig2}c) the pattern remains uniform and defect-free.
For increasing values of $N$, defects akin to those experimentally observed in cholesterics emerge  
\cite{lopez,zumer,zumer3,zumer4,orlova}. They are caused either by the conflict anchoring at the droplet boundary or by the marked modification of the liquid crystal orientation in the bulk. More specifically, in 2d defects can be broadly classified into three groups, namely $\lambda^{\pm m}$, $\tau^{\pm m}$ (where $m$ is the topological charge taking values $\pm 1/2$ and $\pm 1$) and twist disclinations of charge $-1/2$ \cite{EXP7,zumer,fadda}.

If $N=2$, an example of a $\lambda^{+1}$ defect is given in Fig.~\ref{fig2}d where tangential anchoring is set. Here the director arranges in a quasi-2d circular fashion gradually escaping into the third dimension in the middle of the droplet, a structure similar to the one observed in axial drops \cite{lopez,zumer4}. Note that, unlike the defects in nematics, here  it is the local cholesteric pitch axis, rather than the director field, that winds around the defect and is discontinuous at it. The value of the topological charge is the one required to satisfy its global conservation within the droplet \cite{lavrentovich}.
Examples of $\tau$ defects and twist disclinations are shown in Fig.\ref{fig2}e, where perpendicular anchoring is set at the droplet interface. 
Two $\tau^{+1/2}$ defects (blue circles)  appear along the equator due to conflict orientation of the liquid crystal at the edge and in the bulk, while two twist disclinations of charge $-1/2$  (grey circles) are located on opposite sides along the vertical direction. The director of these twist disclinations shows a slight splay in the $y-z$ plane at the top of the defect (where it is actually pinned at the droplet interface), while it twists around the $x$ axis at the bottom \cite{EXP7}. Such defects are connected by a stretched $\lambda^{+1}$ disclination which separates two mirroring splay-bend distortions located symmetrically with respect to the $\lambda$ defect. Once again, the global topological charge is $+1$, as required by these boundary conditions. Note incidentally that,
since in $\tau$ defects the director is singular,
the orientational order drops at their core, thus they can be easily tracked numerically (computing the order at each point).
On the contrary, $\lambda$ lines are more difficult to locate exactly on the lattice since, for reasons of computational efficiency, only a limited number of lattice points can be used to correctly resolve the size of their core, whose radius is 
comparable to the helix pitch $p_0$.
Hence, we prefer referring to 
a ``+1 charged region'' the one containing either a $\lambda^{+1}$ defect or a couple of $\lambda^{+1/2}$ defects, the latter ones often found to merge  (especially in the presence of an external electric field \cite{fadda}). 

Finally, in the absence of anchoring, no defects arise (the global topological charge is zero, see Fig.~\ref{fig2}f) and the director relaxes towards a state exhibiting an almost full-bend arrangement along the vertical direction linked to two symmetric splay-bend distortions in the bulk. 

If $N=4$, a more complex arrangement of the director field is observed. For tangential anchoring (Fig.~\ref{fig2}g), for example, three stretched $\lambda^{+1}$ regions connect couples of defects of charge $-1/2$ localized symmetrically near the interface. With respect to the $N=2$ case, a 2d twisted pattern of the director field, resembling the bipolar cholesteric structure observed, for example, in \cite{zumer,xu_pre}, forms in the bulk of the droplet. For  perpendicular anchoring (Fig.~\ref{fig2}h) one has, once again, three $\lambda^{+1}$ regions linking couples of twist disclinations of charge $-1/2$ plus two $\tau^{+1/2}$ defects located along the equator, a structure partially comparable to that observed in short-pitch cholesteric drops reported in Ref.\cite{orlova}. Such defects disappear if the interface anchoring is absent (Fig.~\ref{fig2}i).  

\subsection{Couples of nematic droplets subject to an oscillatory electric field ($N=0$)}

In this section we focus on the dynamics of couples of drops of liquid crystal hosted in an isotropic fluid and in the presence of an oscillatory electric field. 
The two droplets are accommodated in the middle of the lattice at a distance $d\simeq 2.5R$ between their centers of mass and are relaxed towards their equilibrium state (the ones described in the previous section). The value of such distance is sufficient to minimize contacts between the interfaces of the equilibrated droplets and to prevent coalescence. If the drops come into close contact, their merging is significantly reduced by including a surfactant solute. This is initially set to a constant value uniformly on the lattice and, after relaxation, it accumulates at the droplet interfaces.

Once the equilibrium is attained, a rotating electric field is switched on. The functional form of the components of the field is given by
\begin{equation}
E_{y}=-\frac{\Delta V}{L_y} \sin(\omega t), \hspace{1cm} E_{z}=\frac{\Delta V}{L_z}\cos(\omega t),
\end{equation}
causing a counter-clockwise rotation of the droplets. Here $\Delta V$ is the applied potential, $\omega$ is the frequency of the field and $t$ is the simulation time. The potential $\Delta V$ ranges between $4$ and $10$, since if $\Delta V <4$ the field is too weak to produce a substantial modification of the equilibrium orientation of the liquid crystal, while if $\Delta V>10$ the cholesteric phase turns to a nematic one where the director is aligned along the direction of the electric field. As described in Section II, the values of $\omega$ are varied between $10^{-5}$ (low frequency regime) and $10^{-1}$ (high frequency regime). 

\begin{figure*}
\includegraphics[width=1.\linewidth]{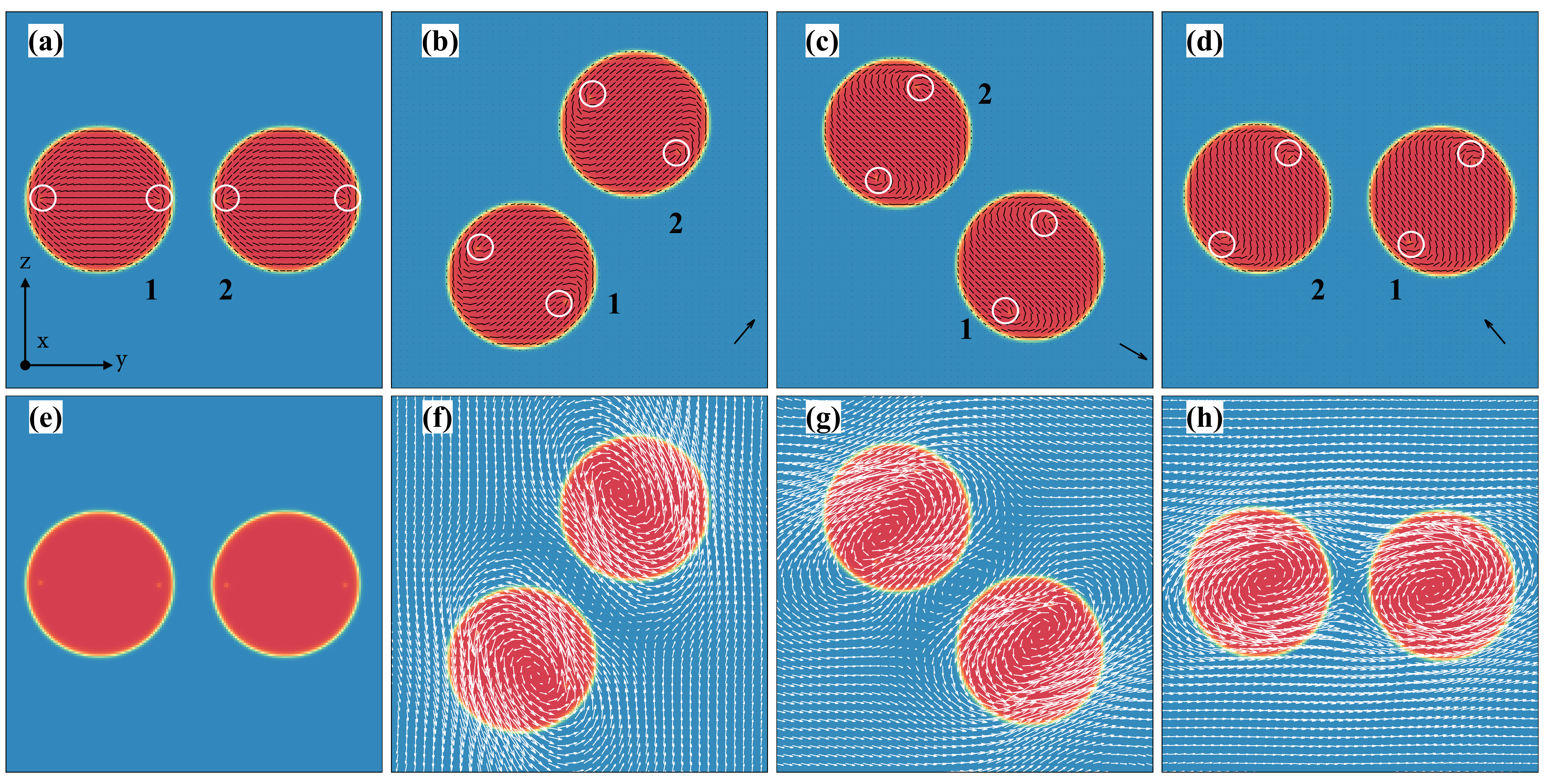}
\caption{Couple of nematic droplets with tangential anchoring ($W=0.04$), $\Delta V\simeq 8$ and $\omega=2 \times 10^{-2}$. The oscillatory field, besides fostering a rotation of the director field within each drop, triggers a counterclockwise rotation of both droplets (identified by numbers $1$ and $2$) around an axis (b,c,d).  In each configuration the director field is aligned with the instantaneous direction of the electric field, indicated by a black arrow. Defects  topological charge $1/2$ (highlighted by white circles)
rotate counterclockwise and essentially remain near the fluid interface. Their motion is sustained by two separate fluid recirculations (e-f-g-h) which affect the local orientation of the liquid crystal and drag the defects. Simulation times are $t=4\times10^{5}$ (a,e), $t=7\times10^{5}$ ((b,f), $t=1.2\times10^{6}$ ((c,g) and $t=1.46\times10^{6}$ (d,h). The color map is the one of Fig.\ref{fig2}a-c.}  
\label{fig3}
\end{figure*}

As benchmark case, we start from two nematic liquid crystal drops with tangential anchoring ($W=0.04$) subject to an oscillatory electric field with $\Delta V\simeq 8$ and frequency $\omega=2 \times 10^{-2}$ (see Fig.~\ref{fig3} and Movie SM1). 
The droplets are marked with numbers $1$ and $2$ and are originally placed horizontally in the middle of the lattice. Once the electric field is turned on, the director rotates counterclockwise remaining almost everywhere aligned with the field, except where two $+1/2$ defects appear (white circles in Fig.~\ref{fig3}a-d), which rotate essentially following a circular trajectory close to the interface.
Such motion is sustained by two separate vortices triggering a counterclockwise rotation of the fluid located within the drops. The vortices 
also modify the velocity field in the surroundings of the drops,  where intense streams promote a counterclockwise rotation of both droplets around an axis located in intermediate fluid film. This effect, in particular, has a purely hydrodynamic nature since it is essentially caused by the momentum transfer from the fast rotating liquid crystal to the droplets and mediated by the fluid located in the middle.
\begin{figure*}
\includegraphics[width=1.\linewidth]{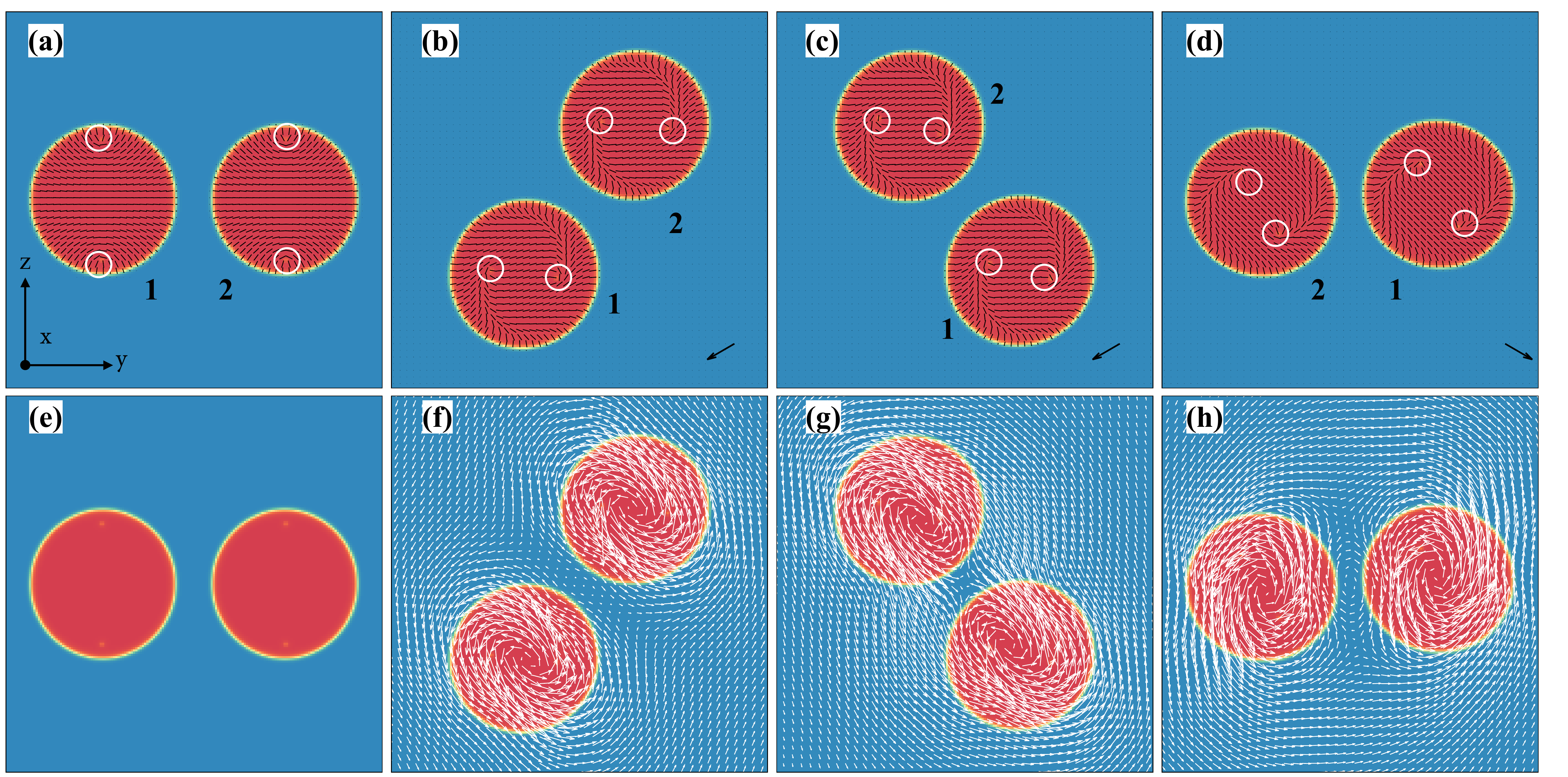}
\caption{Couple of nematic droplets with homeotropic anchoring ($W=-0.04$), $\Delta V\simeq 8$ and $\omega=3 \times 10^{-2}$. The top row (a-d) shows the time evolution of the director field and the defects in a half turn of both droplets (marked with numbers $1$ and $2$) around an axis located in the intermediate fluid film (the blue region). Here defects of charge $-1/2$ rotate counterclockwise moving towards the bulk of the drops. The bottom row (e-h) shows the corresponding velocity field. The black arrow indicates the instantaneous direction of the applied field. Simulation times are $t=4\times10^{5}$ (a,e), $t=7\times10^{5}$ (b,f), $t=10^{6}$ (c,g) and $t=1.22\times10^{6}$ (d,h). The color map is the one of Fig.\ref{fig2}a-c.}
\label{fig4}
\end{figure*}

If the interface anchoring is perpendicular a similar dynamic behavior is observed, once again triggered by two separate fluid vortices  generated by the repeated change of orientation of the liquid crystal (see Fig.~\ref{fig4} and movie SM2, where $\omega=3\times 10^{-2}$). Unlike the previous case, here the topological defects, while rotating, move towards the bulk of the drops, a region where larger distortions occur due to the conflict orientation of the liquid crystal with respect to the one imposed at the droplet interface. 

These two examples clearly suggest that the application of an external oscillatory electric field induces a complex dynamics, consisting of a rotation of each separate drop alongside a circular motion of both drops around each other, plus a persistent rotation of the liquid crystal confined within. To quantitatively evaluate the dynamics of the drops, we compute their angular velocity $\boldsymbol{\omega}^{*}=\frac{\int dV \phi \boldsymbol{r} \times \boldsymbol{u}}{\int dV r^{2}\phi}$  and the angular velocity of the axis joining their centers of mass, defined as 
$\omega^{**}=d \theta/dt$. Here $r$ is the distance from the center of mass of each drop and $\theta$ is the angle the axis forms with the horizontal direction (i.e. the y-axis).
\begin{figure*}
   \begin{minipage}{0.48\textwidth}
     \centering
     \includegraphics[width=.8\linewidth]{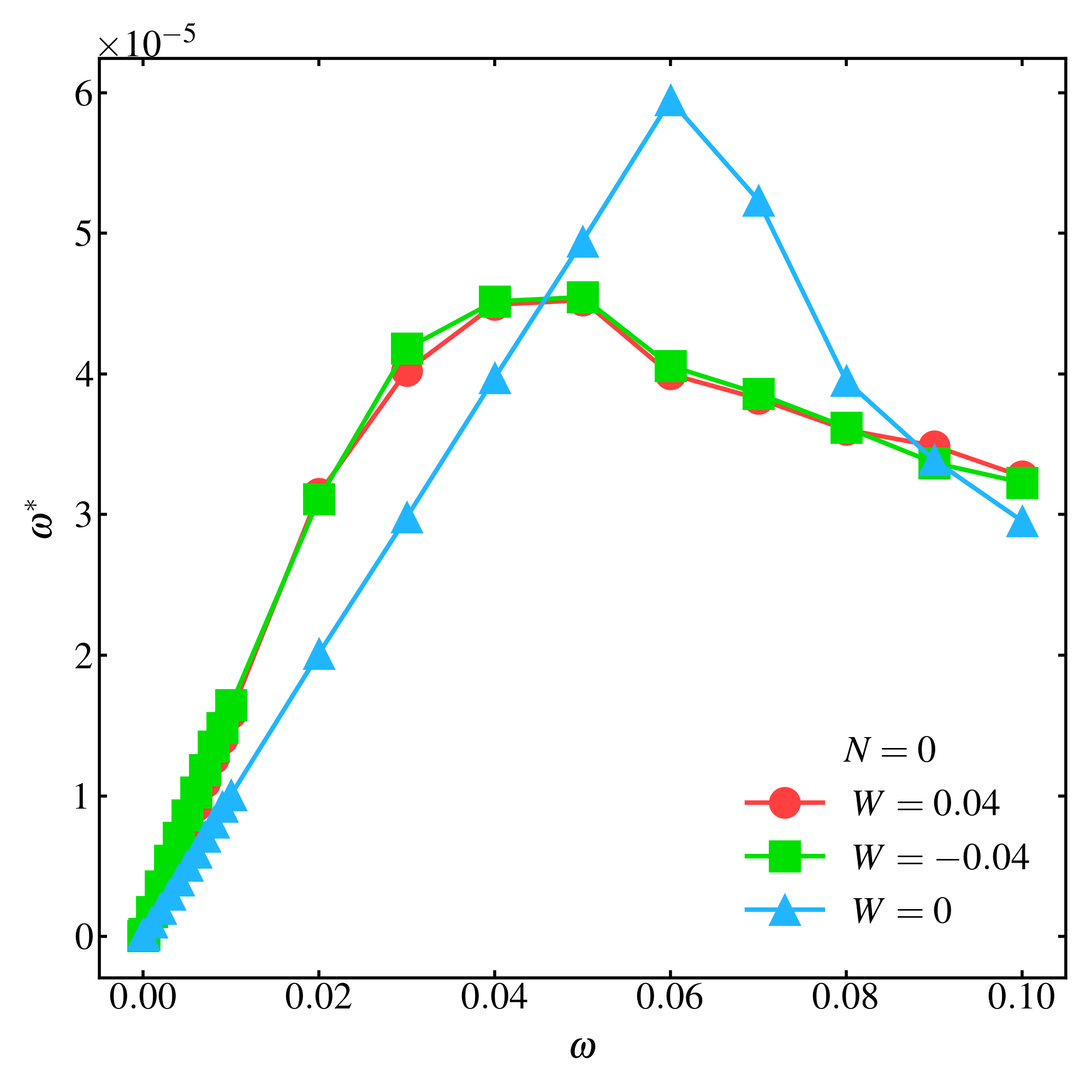}
     \caption{Angular velocity $\omega^{*}$ of each droplet as function of the frequency $\omega$ of the applied field for $N=0$ and anchoring $W=0.04$ (red circle), $W=-0.04$ (green square) and $W=0$ (blue triangle).}\label{NEM_omega}
   \end{minipage}\hfill
   \begin{minipage}{0.48\textwidth}
     \centering
     \includegraphics[width=.8\linewidth]{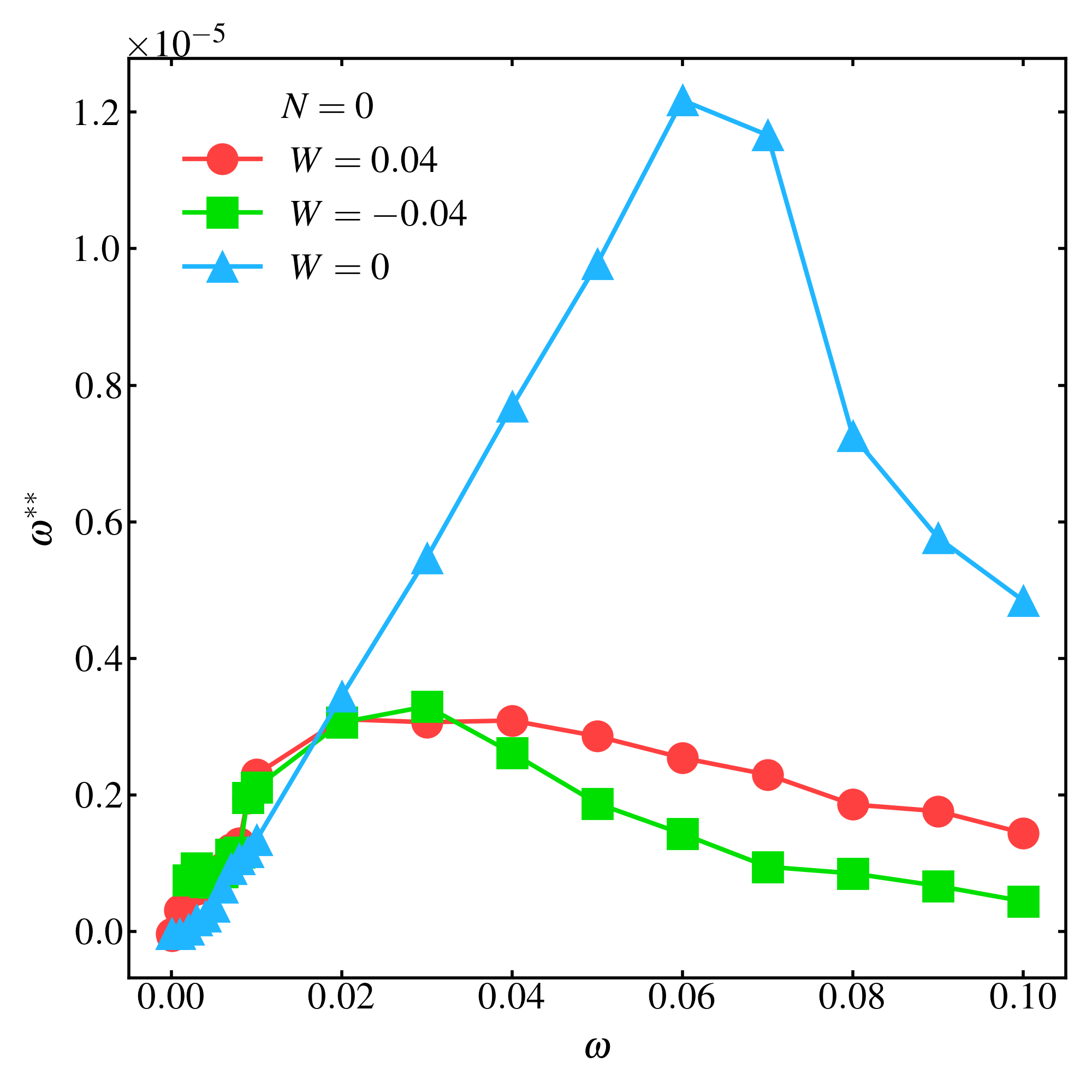}
     \caption{Angular velocity $\omega^{**}$ of the axis joining the centers of mass of the two droplets as function of the frequency $\omega$ of the applied field for $N=0$ and anchoring $W=0.04$ (red circle), $W=-0.04$ (green square) and $W=0$ (blue triangle).}\label{Fig:Data2}
     \label{NEM_omega2}
   \end{minipage}
\end{figure*}
In Fig.~\ref{NEM_omega} and Fig.~\ref{NEM_omega2} we plot $\omega^*$ and $\omega^{**}$ (measured over a time interval of approximately $3.5\times 10^5$ times steps) as a function of the frequency $\omega$ of the applied field and for three different anchoring conditions. For increasing values of $\omega$, the angular speed $\omega^*$ augments almost linearly, essentially regardless of the anchoring at the interface. This behavior lasts as long as $\omega$ is approximately less than $0.05$ whereas, for higher values, $\omega^*$ displays a slight decrease, an effect due to a robust counter-rotating vortex emerging in the fluid between the drops (see section \ref{fluid}). Note, in particular, that the values of $\omega^*$ are about three orders of magnitude lower than those of $\omega$, an indication that a substantial motion of the droplets can be achieved using high frequency fields. Indeed, these ones trigger a fast rotation of the liquid crystal which, in turn, favours the formation of an intense fluid recirculation capable of generating the rotation of the droplet. The axis joining the centers of mass is generally found to rotate at a lower angular speed than that of each droplet (i.e. $\omega^{**}$ is smaller than $\omega^*$), although it is faster for the free-anchoring case. This occurs because the topological defects modify the local orientation of the director, an effect that slightly alters the structure of the velocity field and slows down the speed of rotation of the axis.

It is finally worth noting that the dynamic behavior of the topological defects in a droplet essentially mirrors that in the other one. This is not surprising, since defects of equal topological charge subject to the same applied field are expected to display analogous dynamical features. 
However, unlike nematics, in cholesteric drops defects of different classes come into play,
considerably affecting the response under an applied field. The next section is precisely dedicated to investigating these systems.

\subsection{Cholesterics (N=2)}

\begin{figure*}
\includegraphics[width=1.\linewidth]{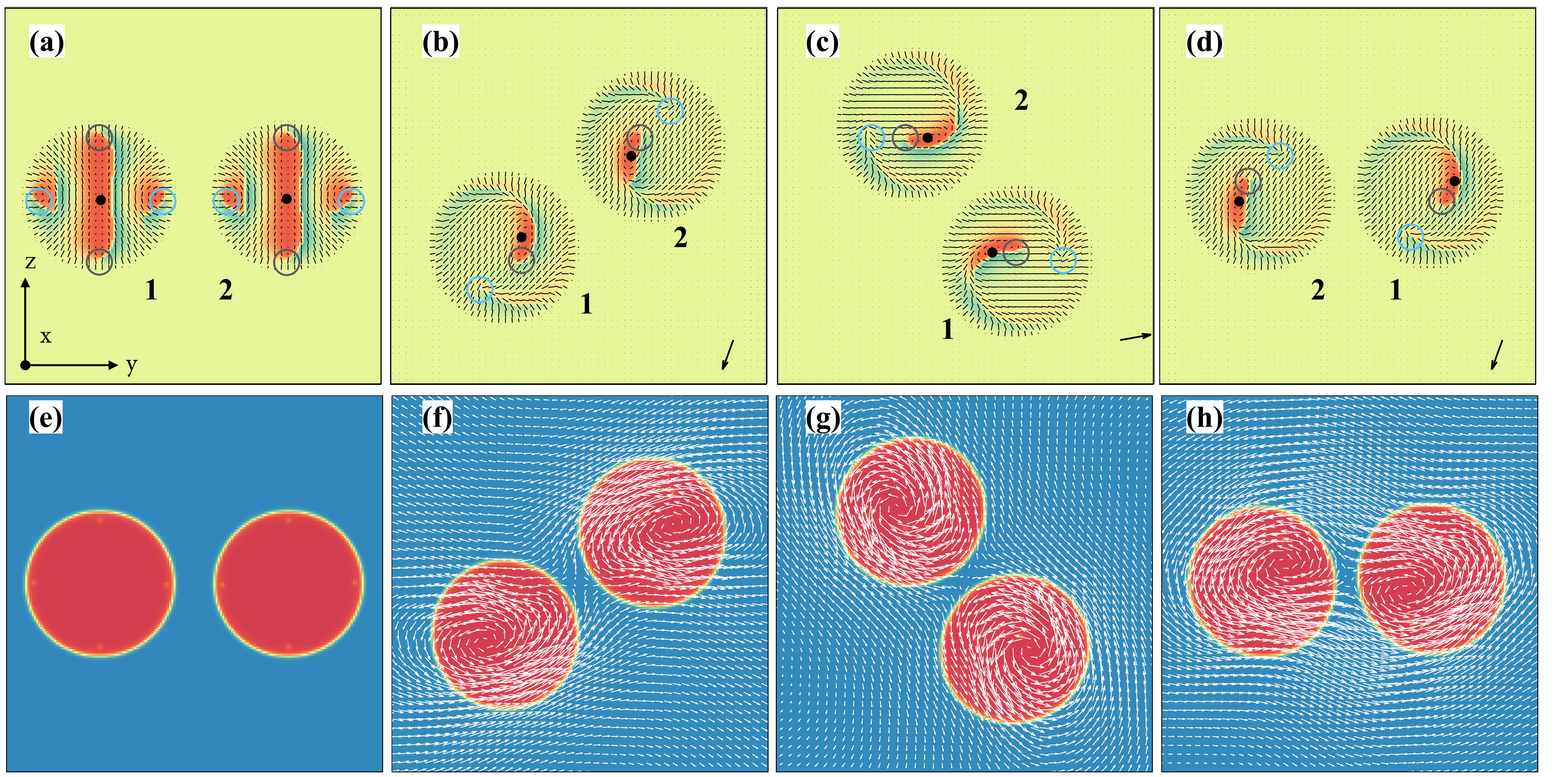}
\caption{Couple of cholesteric droplets in isotropic fluid for $N=2$, $W=-0.04$, $\Delta V\simeq 8$ and $\omega=2\times10^{-2}$.
Once again, under the oscillatory field, liquid crystal and droplets rotate counterclockwise. 
During the rotation (a-d) sustained by two vortices (e-g), a $\tau$ defects of charge $1/2$ and a twist disclinations of charge $-1/2$ (highlighted by blue and grey circles, respectively) annihilate each other, while the remaining two move circularly linked to a stretched $\lambda^{+1}$ region (marked with a black spot in its center). The black arrow at the bottom right indicates the instantaneous direction of the applied field. The snapshots are taken at $t=4\times10^{5}$ (a,e), $t=8\times10^{5}$ (b,f), $t=1.4\times10^{6}$ (c,g) and $t=1.7\times10^{6}$ (d,h). The color map of (a)-(d) is the one of Fig.\ref{fig2}d-i while the color map of (e)-(h) is the  the one of Fig.\ref{fig2}a-c.}
\label{fig6}
\end{figure*}

Here we consider the dynamic response of two cholesteric droplets subject to an oscillatory electric field. We discuss the case in which drops have $N=2$ twists of the director and homeotropic anchoring ($W<0$) is set on their interface (see Fig.~\ref{fig6} and movie SM3). 
In Fig.~\ref{fig6}a the equilibrated droplets show two $\tau^{+1/2}$ defects (blue circles) along the equatorial line ($y$ direction) and two twist disclinations of charge $-1/2$ (grey circles) along the vertical line ($z$-direction), located on opposite sides of a central +1 charged region. Once the field is switched on, both director and defects acquire motion and rotate counterclockwise  (Fig.~\ref{fig6}b-d). This behavior is triggered by two vortices which, in turn, foster the concurrent rotation of the fluid within the drops and of the drops around each other, akin to that observed in the nematic case. However, here only two defects of opposite charge (a twist disclination of charge $-1/2$ and a $\tau^{+1/2}$) survive and move towards the bulk of the droplets, while the other two annihilate each other during their motion. This occurs because 
the droplet is initialized in a metastable state (Fig.~\ref{fig6}a) having an excess of elastic energy provided by the additional defects, subsequently eliminated by the application of the oscillatory electric field. 
Such dynamics occurs in both drops (which basically mirror each other) and persists as long as the field is on (Fig.~\ref{fig6}c-d and g-h). The behavior is considerably simpler if tangential anchoring ($W>0$) is set at the interface. Alongside the orbital motion, within each drop the director aligns with the applied field and rotates as well, bending near the interface to comply with the anchoring conditions. This fosters the concurrent rotation of a bend $\lambda^{+1}$ region (initially located at the center of each drop, see Fig.\ref{fig2}d) which, under field, attains a stretched configuration.

\begin{figure*}
   \begin{minipage}{0.48\textwidth}
     \centering
     \includegraphics[width=.8\linewidth]{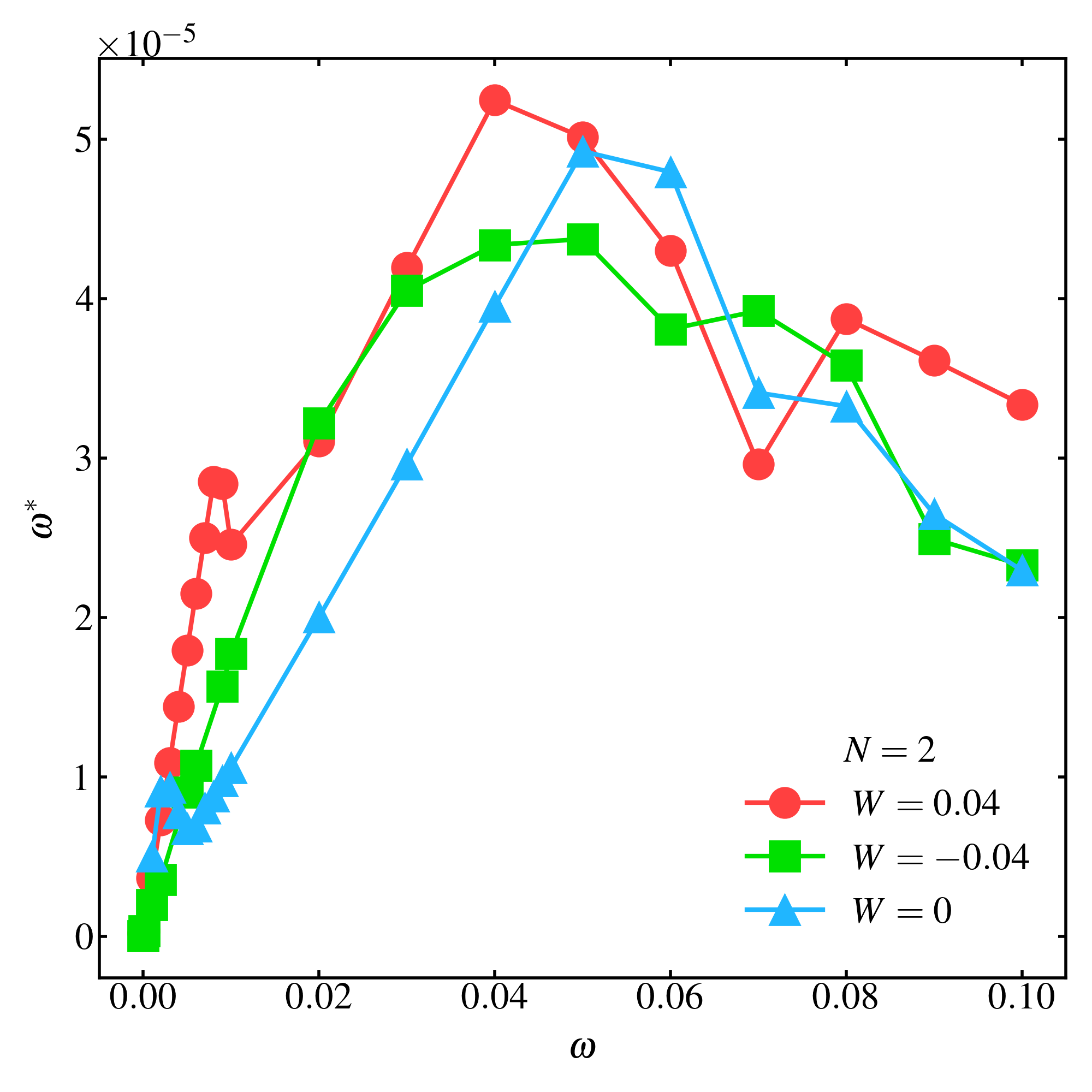}
     \caption{Angular velocity $\omega^{*}$ of each rotating droplet as function of the frequency $\omega$ of the applied field for $N=2$ and anchoring $W=0.04$ (red circle), $W=-0.04$ (green square) and $W=0$ (blue triangle).}\label{Pi_32_omega}
   \end{minipage}\hfill
   \begin{minipage}{0.48\textwidth}
     \centering
     \includegraphics[width=.8\linewidth]{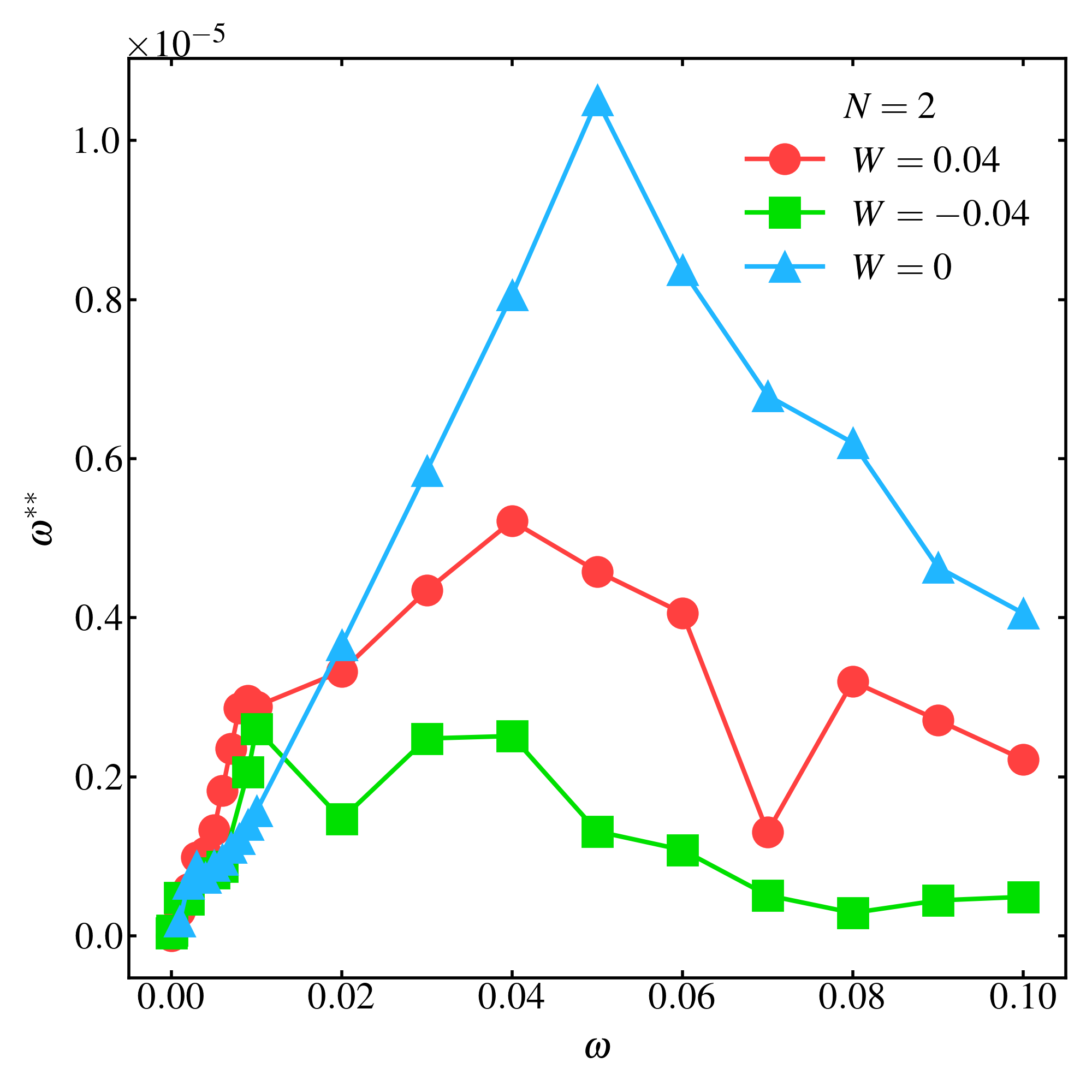}
     \caption{Angular velocity $\omega^{**}$ of the two droplets as function of the frequency $\omega$ of the applied field for $N=2$ and anchoring $W=0.04$ (red circle), $W=-0.04$ (green square) and $W=0$ (blue triangle).}\label{Fig:Data2bis}
     \label{Pi_32_omega2}
   \end{minipage}
\end{figure*}

In Fig.~\ref{Pi_32_omega} and Fig.~\ref{Pi_32_omega2} we show the plots of the angular velocities $\omega^{*}$ and $\omega^{**}$ as function of the frequency $\omega$ of the applied field for three different interface anchoring conditions. The former shows features akin to the nematic case where, for various $W$, an approximate linear growth for $\omega\leq 0.05$ is followed by a mild decrease for higher frequencies, with $\omega^*$ much smaller than $\omega$. On the contrary, the angular speed $\omega^{**}$ of the axis joining the centers of mass of drops with homeotropic anchoring is lower than that of drops with tangential anchoring. This is because the capability of the velocity field to modify the local orientation of the liquid crystal is considerably reduced in the vicinity of the defects, an effect significantly mitigated in drops with tangential anchoring 
(see Fig.~\ref{fig2}d), since the director is continuous near the $\lambda$ defect (see Fig.~\ref{fig2}d). Note finally that $\omega^{**}$ is higher for $W=0$, since here topological defects are absent. 

\subsection{Cholesterics (N=4)}

\begin{figure*}
\includegraphics[width=1.\linewidth]{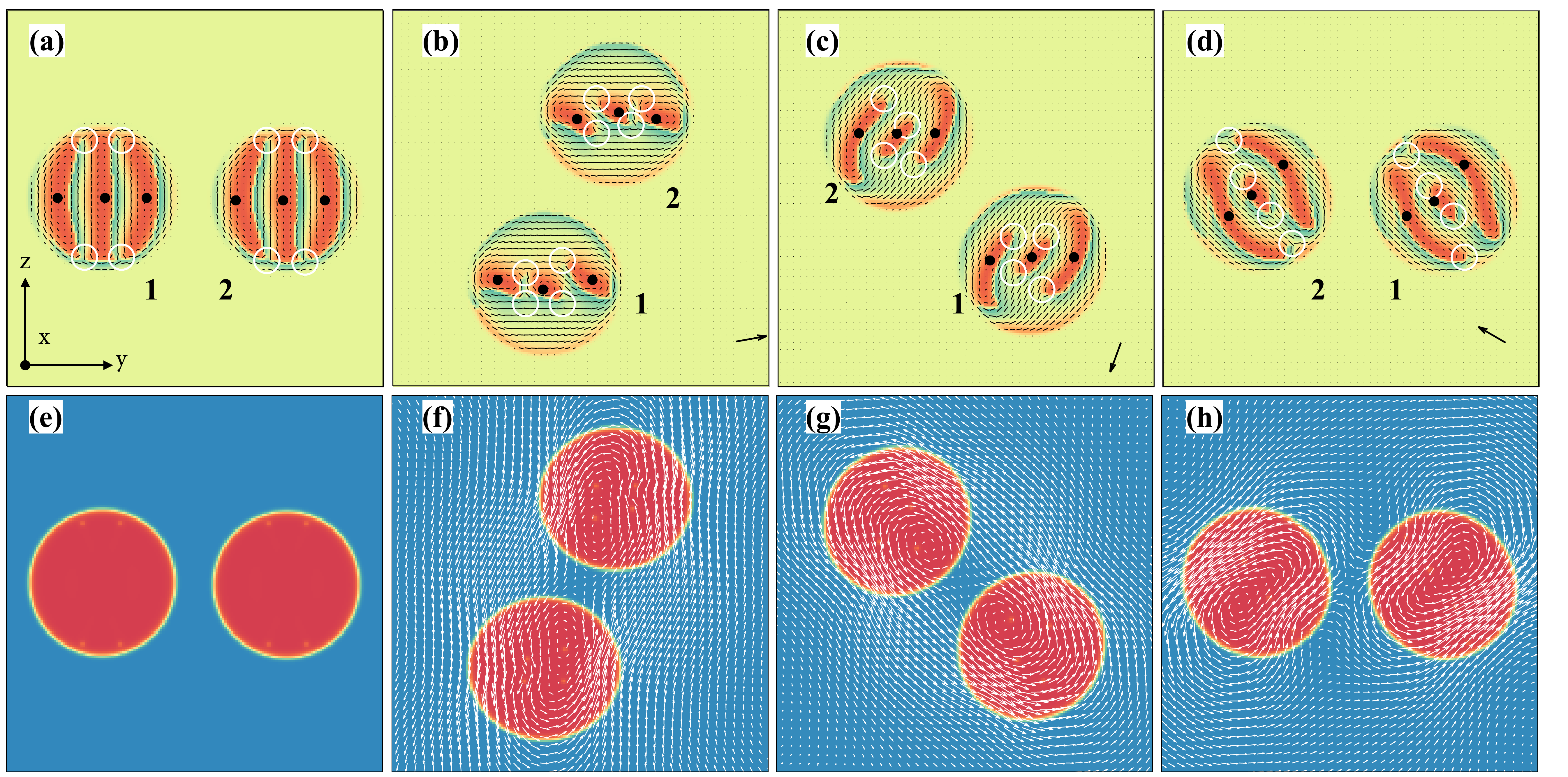}
\caption{Couple of cholesteric droplets in isotropic fluid for $N=4$, $W=0.04$, $\Delta V\simeq 8$ and $\omega=10^{-2}$. Once the field is turned on, four defects of charge $-1/2$ (white circles, a) initially accumulate in the bulk of each drop (b,c) remaining linked to $\lambda$ regions (marked with black spots). Afterwards, they temporarily align essentially mirroring their position within each drop (d). Snapshots are taken at $t=4\times10^{5}$ (a,e), $t=10^{6}$ (b,f), $t=1.6\times10^{6}$ (c,g) and $t=1.95\times10^{6}$ (d,h). The bottom row (e-h) shows the velocity field. The color map of (a)-(d) is the one of Fig.\ref{fig2}d-i while the color map of (e)-(h) is the  the one of Fig.\ref{fig2}a-c.}
\label{fig7}
\end{figure*}

\begin{figure*}
\includegraphics[width=1.\linewidth]{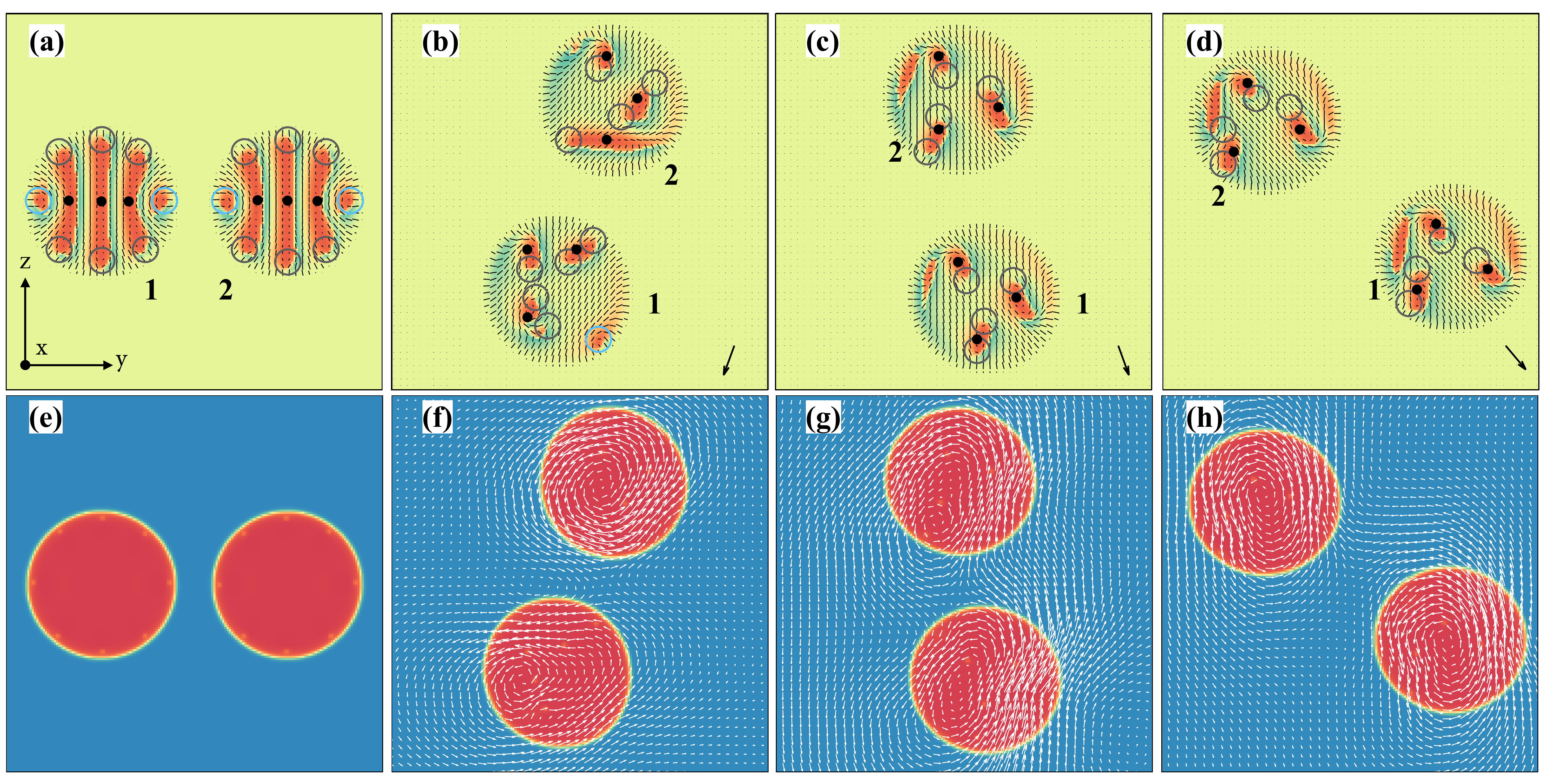}
\caption{Couple of cholesteric droplets in isotropic fluid for $N=4$, $W=-0.04$, $\Delta V\simeq 8$ and $\omega=10^{-2}$. During the rotation, the two $\tau$ defects (circled in blue) annihilate with two twist disclinations (circled in grey), thus only  four twist disclinations survive over longer periods of time (b,c,d). They follow similar trajectories in both drops and remain firmly linked to $\lambda$ regions (marked with black spots). Snapshots are taken at $t=4\times10^{5}$ (a,e), $t=1.6\times10^{6}$ (b,f), $t=2\times10^{6}$ (c,g) and $t=2.902 \times10^{6}$ (d,h). The bottom row (e-h) shows the velocity field. The color map of (a)-(d) is the one of Fig.\ref{fig2}d-i while the color map of (e)-(h) is the  the one of Fig.\ref{fig2}a-c.}
\label{fig8}
\end{figure*}

A considerably more complex dynamics is observed when $N=4$. In Fig.~\ref{fig7} and Fig.~\ref{fig8} we show a sequence of  configurations of two cholesteric drops with tangential and homeotropic anchoring subject to a rotating electric field (see also movie SM4 and SM5). 

In the former, the two equilibrated drops, accommodated symmetrically with respect to the $z$-axis, show four $-1/2$ defects (white circles) located near the interface and connected to three internal \lq\lq +1 charged regions" (Fig.~\ref{fig7}a-e). Once the field is turned on, both drops are set in motion following a trajectory essentially akin to the cases aforementioned. However, the defect dynamics displays fully distinctive features. The four $-1/2$ defects rotate and move towards the bulk of the droplets, while the three $\lambda$ defects initially shorten, attaining an almost spot-like configuration (Fig.~\ref{fig7}b-f), and afterwards stretch (Fig.~\ref{fig7}c-g) exhibiting pronounced bends induced by the curvature of the interface. Note, in particular, that the typical mirror dynamic behavior of the defects in both drops is temporarily lost and partially restored later on (Fig.~\ref{fig7}d). Here the four $-1/2$ defects arrange momentarily in a single file, two at the center of the drop linked by a spot-like $\lambda$ defect and the remaining two, near the interface, connected by a couple of elongated $\lambda$ regions. 

If the interface anchoring is homeotropic, at equilibrium each droplet displays two $\tau^{1/2}$ defects (blue circles) located along the equatorial line and six twist disclinations, (grey circles) placed near opposite parts of the interface and linked by three elongated $\lambda$ regions (see Fig.~\ref{fig8}a). The application of the oscillatory electric field, besides fostering the typical dynamics of the drops overall akin to the previous cases (see Fig.~\ref{fig8}b-d and f-h), also induces annihilation among defects of opposite charges, thus diminishing their number. Indeed one observes a temporary state in which 
three $\lambda$ coexist either with four twist disclinations (Fig.~\ref{fig8}b, drop 2) or with five twist disclinations plus a single $\tau^{1/2}$ (Fig.~\ref{fig8}b, drop 1)  which, later on, annihilates with a twist disclination of charge $-1/2$ (Fig.~\ref{fig8}c). This leaves both droplets with three $\lambda$ and four twist disclinations (Fig.~\ref{fig8}b), i.e. the minimum number of defects required to preserve the topological charge in a droplet with homeotropic anchoring and $N=4$ twists. 
Note also that the rotational motion of both drops around an axis, located in the film of fluid, arrests. This occurs because the non-uniform orientation of the liquid crystal favours a temporary shift of the fluid vortexes off center (see, for instance, Fig.~\ref{fig8}g), an effect promoting the formation of repulsive fluid flows located within the film separating the droplets. Afterwards, the vortices regain their approximately central position (Fig.~\ref{fig8}h), considerably weakening the momentum transfer between the drops thus hindering any further rotation.

\begin{figure*}
\begin{minipage}{0.48\textwidth}
\centering
\includegraphics[width=.8\linewidth]{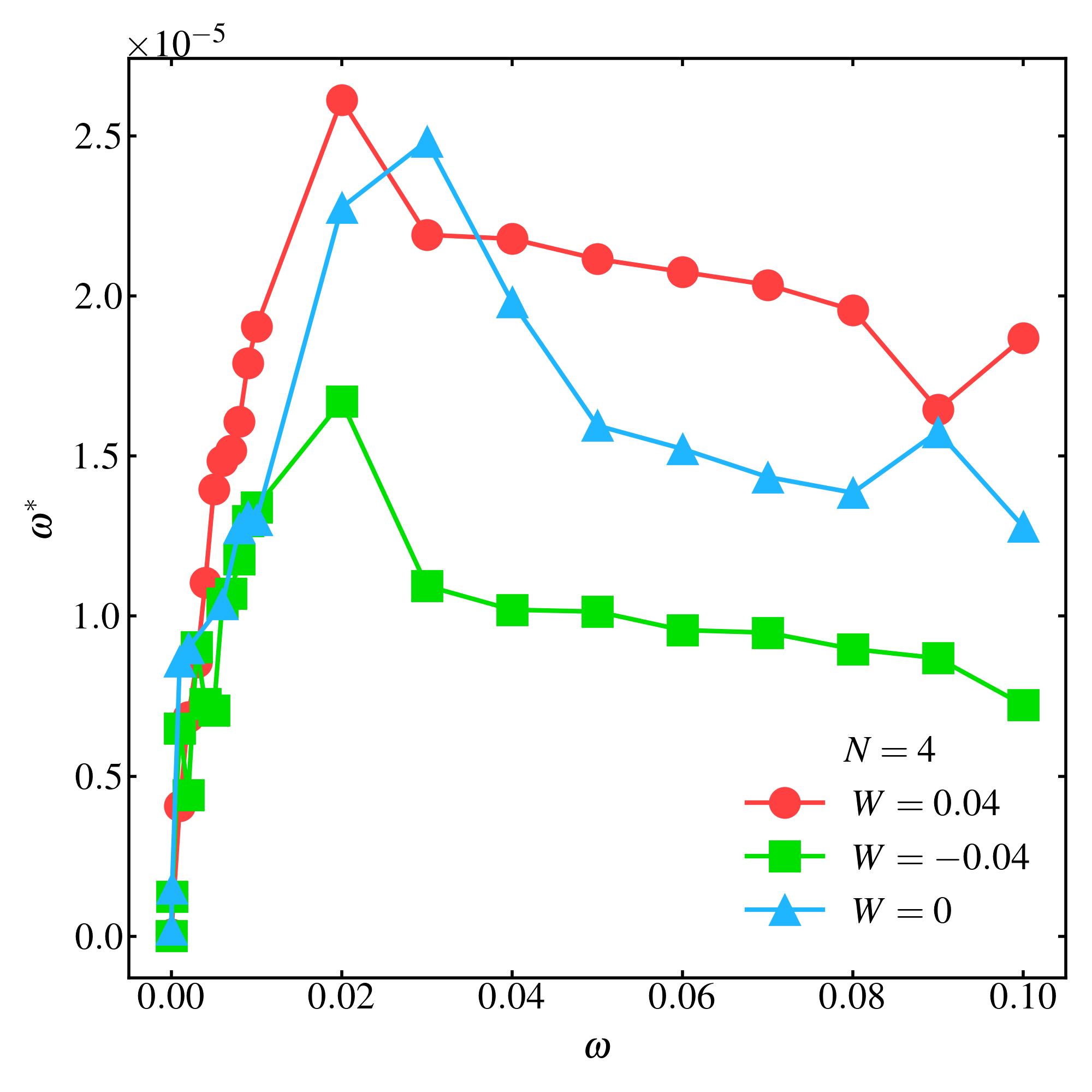}
\caption{Angular velocity $\omega^{*}$ of each rotating droplet as function of the frequency $\omega$ of the applied field for $N=4$ and anchoring $W=0.04$ (red circle), $W=-0.04$ (green square) and $W=0$ (blue triangle).}\label{Pi_16_omega}
\end{minipage}\hfill
\begin{minipage}{0.48\textwidth}
\centering
\includegraphics[width=.8\linewidth]{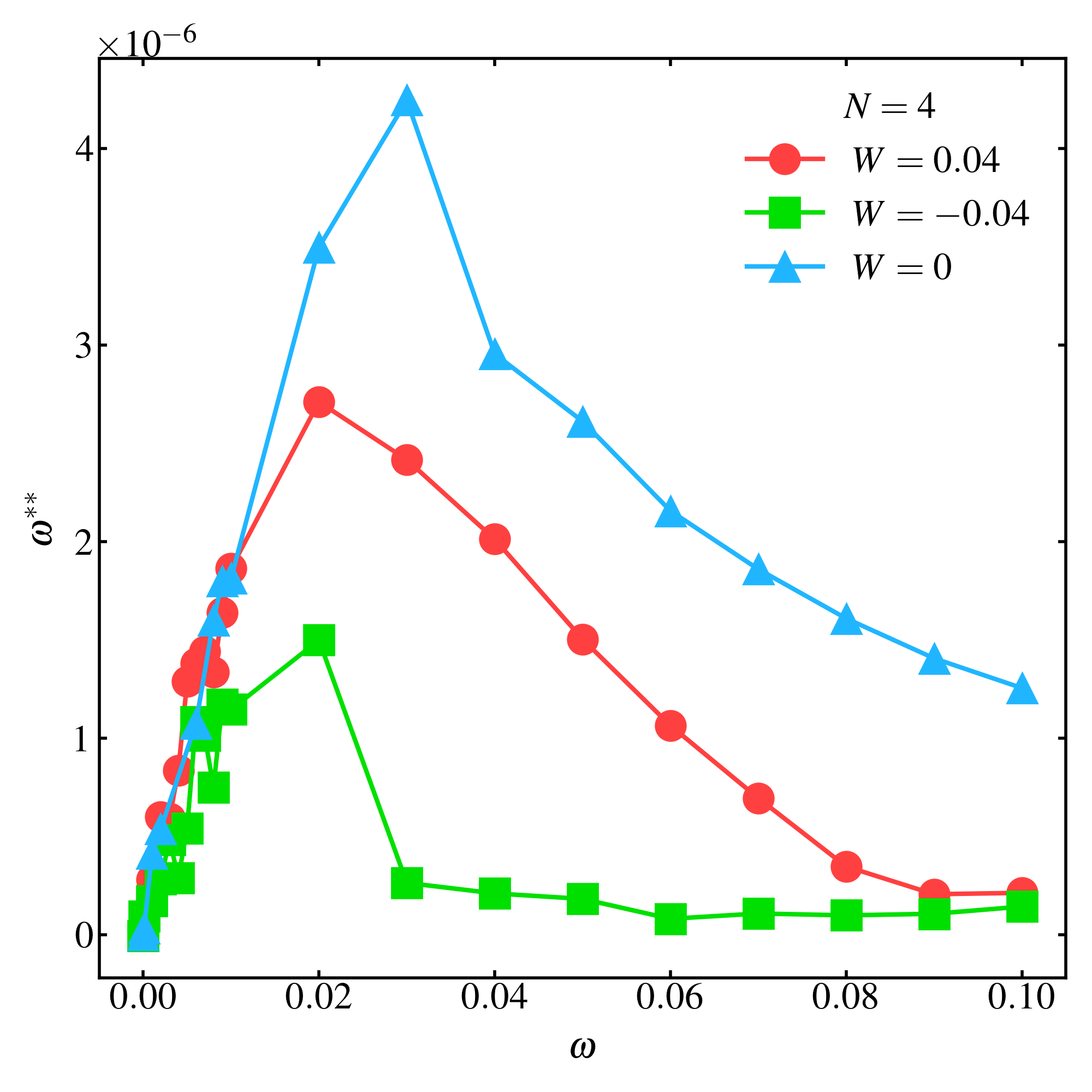}
\caption{Angular velocity $\omega^{**}$ of the two droplets as function of the frequency $\omega$ of the applied field for $N=4$ and anchoring $W=0.04$ (red circle), $W=-0.04$ (green square) and $W=0$ (blue triangle).}\label{Pi_16_omega2}
\end{minipage}
\end{figure*}

In  Fig.~\ref{Pi_16_omega}  and  Fig.~\ref{Pi_16_omega2},  we  show  the  plots  of $\omega^*$ and $\omega^{**}$,  which share analogous features with the ones obtained for $N=2$. The values of $\omega^*$ and $\omega^{**}$, for example, are once again significantly lower in the homeotropic anchoring case, because of the larger resistance opposed by the liquid crystal to change its orientation. Also, the usual linear growth observed for $\omega<0.02$ is followed, at higher frequencies, either by a short decrease quickly stabilized to constant values for $\omega^*$ (Fig.~\ref{Pi_16_omega}) or by a steep descent for $\omega^{**}$ (Fig.~\ref{Pi_16_omega2}). However, here their values are much smaller than the ones computed for $N=2$, an effect due to the larger number of defects produced in cholesterics drops with a higher pitch. 

We finally note that, regardless of nature of the liquid crystal (whether nematic or cholesteric) and interface anchoring, both $\omega^*$ and $\omega^{**}$ show a distinctive feature for $\omega\simeq 0.02$, a value after which they generally diminish either gently (such as Fig.~\ref{fig7}) or more rapidly (as in Fig.~\ref{Pi_16_omega2}). An analogous result has been also experimentally observed, for example, in cholesteric samples subject to a temperature gradient and additionally exposed to an AC field \cite{lehmann2}. Before concluding, we dedicate the next section to clarifying this behavior.

\subsection{Fluid velocity at high frequency}\label{fluid}

\begin{figure*}
\includegraphics[width=0.8\linewidth]{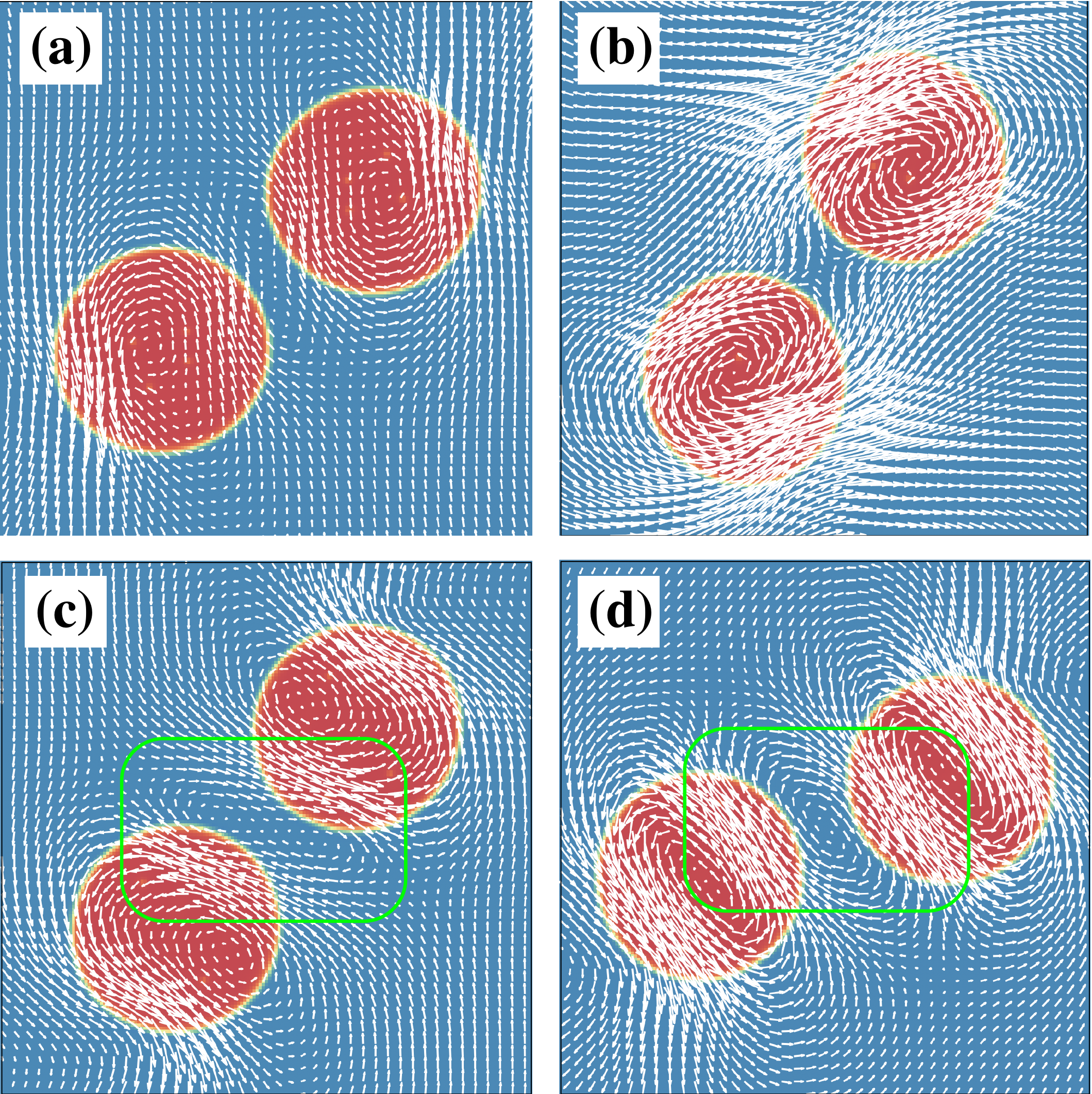}
\caption{Instantaneous configurations, taken at $t=7.5\times 10^{5}$ for different values of $\omega$, of two cholesteric drops with $N=4$ and $W=0.04$. In particular, $\omega$ is equal to $0.01$ (a), $0.02$ (b), $0.03$ (c) and $0.06$ (d). The green rectangle in (c) and (d) highlights a fluid vortex formed within the film separating the drops. The color map is the one of Fig.\ref{fig2}a-c.}
\label{fig13}
\end{figure*}

In Fig.~\ref{fig13} we show, for example, the instantaneous configurations of two cholesteric drops with $N=4$ and $W=0.04$, subject to a rotating electric field for different values of frequency $\omega$. While in (a) and (b) (where $\omega$ is $0.01$ and $0.02$, respectively) the structure of the fluid velocity closely resembles the ones previously observed (i.e. two well-defined vortices located within each drop), in (c) and (d) (where $\omega$ is $0.03$ and $0.06$) a robust counter-rotating vortex emerges in the film of fluid located between the drops. This one essentially results from the combination of opposite branches of the two fluid vortices facing each other. Thus, at high frequency of the applied field, the velocity exhibits three distinct vortices (see Fig.~\ref{fig13}d), two placed within the drops favouring their rotation and a further one located in between hampering the motion. Such effect is amplified for increasing values of $\omega$, since a faster rotating liquid crystal would transfer a larger momentum to the surrounding fluid strengthening the  vortices and thus slowing down the droplets rotation. 

\section{Conclusions}
To summarize, we have numerically studied the dynamics of a couple of liquid crystal droplets (nematic and cholesteric with weak dielectric anisotropy) immersed in an isotropic phase and subject to an oscillatory electric field. We have considered liquid crystals made of rod-like shaped molecules whose orientation (described by the order parameter ${\bf Q}$) couples quadratically to the electric field, a description which holds as long as flexoelectricity (an elastic distortion generated by the polarization induced by the electric field) and ionic effects are weak  \cite{tarasov2003}. Simulations are run using a hybrid LB approach, in which the Navier-Stokes equation for the fluid velocity is integrated using a standard LB algorithm while the equations of the concentration of the surfactant and liquid crystal plus that governing the orientation of the latter are solved adopting a finite difference scheme. 

Drops are selected in terms of the number $N$ of $\pi$ twists of the liquid crystal (i.e. the pitch) as well as the direction of the interface anchoring, either perpendicular or tangential. We have specifically focused on nematic ($N=0$) and cholesterics ($N=2$ and $N=4$) with strong tangential ($W=0.04$) and perpendicular ($W=-0.04$) anchoring. If $N=0$, for example, the liquid crystal preferentially aligns along the direction of the applied field and rotates coherently at an angular speed $\omega$ equivalent to the one imposed by the field. Topological defects of charge $\pm 1/2$ generally follow a circular trajectory, either near the interface (with tangential anchoring) or closer to the center of the drops (with homeotropic anchoring), mirroring each other during the motion. In both cases, the velocity field exhibits two vortices approximately centered within each drop, which trigger the rotation of the fluid confined within as well as that to the drops around each other. This is a fully hydrodynamic effects since it is originated by the momentum transfer of the liquid crystal to the surrounding fluid.  
In the cholesteric phase, further defects emerge due to the combination of an increased number of twists ($N=2$ and $N=4$) of the director field and its conflict orientation with that imposed at the interface anchoring. Under an oscillatory field, $\lambda$ defects periodically elongate and shorten while $\tau$ ones follow a complex trajectory, remaining anchored at the extremities of each $\lambda$ defect. The velocity field shows a structure akin to that observed for the nematic counterpart and it is, once again, capable of triggering the orbital rotation of drops and of the fluid located within. Our results also show that the angular speed at which these two processes occur augments approximately linearly for low values of $\omega$ while, for increasing values, it decreases and then stabilizes to roughly constant values, basically regardless of the nature of the liquid crystal considered. This last behavior is shown to have a purely hydrodynamic origin, since it is caused by the formation, in the middle of the drops, of a further counter-rotating vortex hindering their rotation.

The results discussed in this work show that, alongside magnitude and frequency of the applied field, the motion of the drops is decisively affected by the nature of the topological defects, the pitch and the elasticity of the liquid crystal. However a number of questions remains open. It would be of interest, for example, investigating to which extent the size of the droplets may affect the defect dynamics and the complex rotational motion described in this paper. In addition, diminishing the reciprocal distance may alter the structure of the velocity field in fluid film separating the drops and, in the worst case scenario, favour their merging, especially for strong enough electric fields. An alternative dynamic behavior is also expected to occur if the approximation of strong interface anchoring is released or, even more intriguingly, if the effects of the three bulk elastic constant (splay, twist and bend) are separately considered. Finally, although quasi-2d chiral samples can be experimentally realized \cite{cluzeau}, a more realistic picture could be conveyed by fully three dimensional simulations, where further complex defect patterns can significantly enrich the dynamics under an external field.

\section*{Acknowledgements}
Simulations have been performed at Bari ReCas e-Infrastructure funded by MIUR, Italy through the program PON Research and Competitiveness 2007-2013 Call 254 Action I. F. F. acknowledges funding from the Japan Society for the Promotion of Science (JSPS) KAKENHI grant 17H01083 and funding from the National Science Foundation under Grant No. NSF PHY-1748958 and D-ITP consortium, a program of the Netherlands Organization for Scientific Research (NWO) that is funded by the Dutch Ministry of Education, Culture and Science (OCW). A. T. acknowledges funding from the European Research Council under the European Union's Horizon 2020 Framework Programme (No. FP/2014-2020) ERC Grant Agreement No.739964 (COPMAT). A. T. also warmly thanks Livio Carenza and Davide Marenduzzo for useful discussions. A. L. acknowledges funding from MIUR Project No. PRIN 2020/PFCXPE.
This work was performed under the auspices of GNFM-INdAM.

\bibliography{biblio}

\end{document}